\documentclass[journal]{IEEEtran}
\usepackage{amsmath,amsfonts}
\usepackage{amsthm}
\usepackage{algorithmic}
\usepackage{algorithm}
\usepackage{array}
\usepackage[caption=false,font=normalsize,labelfont=sf,textfont=sf]{subfig}
\usepackage{textcomp}
\usepackage{stfloats}
\usepackage{url}
\usepackage{verbatim}
\usepackage{graphicx}
\usepackage{cite}
\usepackage{booktabs}
\usepackage{makecell}
\usepackage{xcolor}
\usepackage{braket}
\usepackage{soul}
\usepackage{bm}

\renewcommand{\arraystretch}{1.4}
\newtheorem{problem}{Problem}

\begin{document}

\title{Quantum-Based $k$-Coverage Optimization for UAV-Aided Search-and-Rescue Missions}

\author{Halim~Lee, Suhui~Jeong, Na~Young~Kim, and Jiwon~Seo,~\IEEEmembership{Senior Member,~IEEE}
\thanks{This work was supported in part by the National Research Foundation of Korea (NRF), funded by the Korean government (Ministry of Science and ICT, MSIT), under Grants RS-2024-00358298 and RS-2025-25393156; 
in part by the Korea Aerospace Administration (KASA), under Grant RS-2022-NR067078;  
in part by Grant RS-2024-00407003 from the ``Development of Advanced Technology for Terrestrial Radionavigation System'' project, funded by the Ministry of Oceans and Fisheries, Republic of Korea; 
in part by the Unmanned Vehicles Core Technology Research and Development Program through the NRF and the Unmanned Vehicle Advanced Research Center (UVARC), funded by the MSIT, Republic of Korea, under Grant RS-2020-NR046546; 
and in part by the Institute of Information \& Communications Technology Planning \& Evaluation (IITP) through the Information Technology Research Center (ITRC) program, funded by the MSIT, under Grant IITP-2025-RS-2024-00437494.
\textit{(Halim Lee and Suhui Jeong are co-first authors.)
(Corresponding Authors: Na Young Kim, Jiwon Seo.)}}
\thanks{Halim~Lee, Suhui~Jeong, and Jiwon~Seo are with the School of Integrated Technology, Yonsei University, Incheon 21983, Republic of Korea (e-mail: halim.lee@yonsei.ac.kr; ssuhui@yonsei.ac.kr; jiwon.seo@yonsei.ac.kr).}
\thanks{Na~Young~Kim is with the Institute for Quantum Computing, the Department of Electrical and Computer Engineering, and the Waterloo Institute for Nanotechnology, University of Waterloo, Waterloo, ON N2L 3G1, Canada (e-mail: \mbox{nayoung.kim@uwaterloo.ca}).}
\thanks{Copyright (c) 2026 IEEE. Personal use of this material is permitted. However, permission to use this material for any other purposes must be obtained from the IEEE by sending a request to pubs-permissions@ieee.org.
}
}

\markboth{IEEE Internet of Things Journal,~Vol.~00, No.~0, April~2026}%
{Lee \MakeLowercase{\textit{et al.}}: Paper Title}


\maketitle

\begin{abstract}
In large-scale disaster scenarios, rapid localization of missing persons is a critical challenge for search-and-rescue (SAR) operations. Unmanned aerial vehicles (UAVs) equipped with radio frequency (RF) receivers can support RF-based localization by collecting signals emitted from mobile devices at spatially distributed sensing locations. This paper addresses the resulting waypoint-selection problem: determining a minimum set of UAV waypoints that provides at least threefold coverage of every potential target location. We formulate this task as an extended $k$-coverage problem that independently defines the UAV-navigable and target regions, and derive an exact-penalty quadratic unconstrained binary optimization (QUBO) formulation with a sufficient penalty condition that preserves feasibility and minimum waypoint cardinality. The QUBO is mapped to an Ising-form cost Hamiltonian and evaluated using the quantum approximate optimization algorithm (QAOA) on both a noise-free simulator and IBM's 127-qubit Eagle processor. On the tested simulator instances, QAOA recovers the known minimum-cardinality solutions. Across the rectangular hardware test cases, the mean 3-coverage ratio exceeded 95\%. In the campus-scale evaluation, ten hardware executions achieved 99.3\% $\pm$ 2.1\% mean 3-coverage with a 90\% feasible-run rate, while the shortest feasible flight path was up to 37.0\% shorter than those of the deterministic grid-based baselines. Additional comparisons with classical optimization and learning-based baselines are provided, together with computational and quantum-resource analyses for larger generated instances. These results establish an exact QUBO representation for RF-based SAR waypoint selection and characterize its implementation on current gate-based quantum hardware.
\end{abstract}


\section{Introduction}
\IEEEPARstart{L}{arge-scale} disasters, including earthquakes, floods, and storms, impact vast regions and cause extensive damage to both human lives and infrastructure. According to the 2024 Emergency Events Database (EM-DAT), 393 natural hazard-induced disasters in 2024 resulted in 16,753 fatalities, affected 167.2 million people, and caused an estimated economic loss of US\$242 billion \cite{EMDAT24:2025}. In such large-scale disaster scenarios, missing persons are often scattered across vast and complex terrains, making rapid and efficient search efforts extremely challenging. Consequently, search and rescue (SAR) operations demand time-sensitive, autonomous, and scalable strategies capable of covering large areas in real time.

Recently, Internet of Things (IoT)-enabled unmanned aerial vehicles (UAVs) have emerged as a transformative solution for SAR missions. Functioning as aerial sensor nodes, UAVs enhance search capabilities by collecting and transmitting real-time data from the air. Various sensing modalities have been explored, including camera-based \cite{Liu24:Explainable}, thermal imaging-based \cite{Zou23:UAV}, and radar-based \cite{Wang24:An} techniques for detecting missing persons. However, these methods often fail to detect individuals trapped indoors or buried beneath debris, thereby limiting their effectiveness in real-world disaster scenarios.

To overcome this limitation, recent studies \cite{Perazzo16:Drone, Yuan22:A, Moon24:HELPS, Shih17:Doppler, Wang13:Feasibility} have proposed leveraging wireless radio frequency (RF) signals naturally emitted by the mobile devices of missing individuals (hereafter referred to as ``targets'').
Examples include Wi-Fi probe requests \cite{Shih17:Doppler, Wang13:Feasibility} and cellular uplink signals \cite{Moon24:HELPS}, which---although intermittent---can be captured by UAVs equipped with suitable RF receivers.
By hovering at strategically selected waypoints and collecting multiple RF measurements (e.g., time-of-arrival (TOA) or received signal strength (RSS)) from spatially distinct positions, UAVs can perform range-based localization to estimate the positions of targets.

A central challenge in this RF-based localization approach lies in waypoint planning. To enable target localization, the UAV must visit a set of spatially distributed waypoints surrounding each target. However, in time-critical SAR missions, every additional waypoint increases the flight path and mission duration, whereas insufficient coverage may provide too few measurements for target localization.
Therefore, the objective is not merely to identify feasible waypoints but to determine the minimum set of waypoints that guarantees sufficient spatial coverage, commonly referred to as the waypoint optimization problem.

This problem is NP-hard~\cite{Hefeeda07:Randomized}. At the instance sizes accessible to current quantum hardware, mature classical solvers can handle it effectively; in our campus-scale evaluation, for example, an exact integer linear programming (ILP) formulation returns a provably optimal waypoint set in well under a second (see Section~\ref{subsec:OptLearnComparison}). Exact solution nevertheless remains exponential in the worst case, and larger search areas or finer grid resolutions may therefore motivate greater reliance on heuristic or approximation strategies that do not guarantee global optimality~\cite{Hefeeda07:Randomized,Elhoseny18:Optimizing,Tarnaris20:Coverage}. Quantum optimization has been proposed as an alternative computational paradigm for such combinatorial problems~\cite{Au23:NP,Chatterjee24:Solving}, providing motivation to develop and evaluate quantum-compatible formulations for the SAR waypoint-selection problem considered here.

Quadratic unconstrained binary optimization (QUBO) provides a common representation for binary combinatorial optimization that can be addressed by both classical QUBO-based solvers and quantum optimization methods. A QUBO objective can be mapped to an Ising-form cost Hamiltonian and addressed using variational quantum algorithms such as the quantum approximate optimization algorithm (QAOA). As gate-based quantum processors continue to improve in scale and capability, application-specific QUBO formulations provide a means to investigate the potential of quantum optimization while directly characterizing the circuit and hardware resources required for their implementation.

Accordingly, this work develops an exact-penalty QUBO formulation of the SAR waypoint-selection problem and investigates its implementation on current gate-based quantum hardware. We formulate the problem as an extended $k$-coverage problem and establish a sufficient condition under which every global QUBO minimizer corresponds to a feasible minimum-cardinality solution of the original problem. The resulting QUBO is implemented using QAOA on both a noise-free simulator and real quantum hardware. Classical optimization and learning-based baselines, together with computational and quantum-resource analyses, are further used to characterize the performance and resource requirements of the resulting implementation.

To the best of our knowledge, this work presents the first exact-penalty QUBO formulation and gate-based quantum implementation of the extended $k$-coverage UAV waypoint-selection problem for RF-based SAR. The main contributions are summarized as follows:

\begin{itemize}
\item \textbf{Extended $k$-coverage formulation.}
We formulate UAV waypoint selection as an extended $k$-coverage problem that explicitly separates the UAV-navigable region from the target region in which coverage must be guaranteed. This formulation permits candidate waypoints to be located both within and outside the target region and reduces to conventional $k$-coverage when the two regions coincide.

\item \textbf{Exact-penalty QUBO with an exactness guarantee.}
We derive an exact-penalty QUBO formulation using a logarithmic-size binary encoding of the coverage surplus and establish the sufficient condition $\lambda>|\boldsymbol{W}|$, under which every global QUBO minimizer is feasible for the original problem and has minimum waypoint cardinality over the feasible set.

\item \textbf{Real-hardware implementation and comprehensive evaluation.}
We implement the resulting QUBO using QAOA on both a noise-free simulator and a real 127-qubit quantum processor. The evaluation includes classical exact, heuristic, metaheuristic, and learning-based baselines, and sensitivity analyses of the principal QUBO and QAOA settings.

\item \textbf{Computational and quantum-resource analysis.}
We characterize the computational cost of coverage construction, QUBO generation, and QAOA-based optimization, together with the logical circuit width, number of nonzero QUBO coefficients, and circuit depth. We further analyze formulation-level resource growth for fixed-density instances containing up to 1,992 candidate waypoints.
\end{itemize}

\section{Background} \label{Sec:Background}

\subsection{UAV path planning for SAR}
Recent UAV--IoT research has addressed networking, autonomous operation, and planning, including QoS-aware multi-UAV communication~\cite{Chen23:QoS}, learning-based routing~\cite{Wang22:Learning}, energy-constrained data collection~\cite{Fan24:Energy}, reinforcement-learning-based task allocation and conflict-free path planning~\cite{Zhao24:Reinforcement}, and trajectory, communication, and resource management~\cite{Bai23:Toward}. More directly related to emergency target localization, Lee and Seo developed rigidity-based multi-UAV trajectory optimization to improve cooperative sensing geometry~\cite{Lee26:Rigidity}. These studies primarily focus on communication, routing, resource management, task allocation, or trajectory-level planning. In contrast, the present work addresses SAR-oriented sensing-location planning, selecting a compact set of spatially distributed UAV waypoints that provides redundant RF observations over the candidate target region and formulating the resulting waypoint-selection problem as an exact-penalty QUBO for gate-based quantum optimization.

Within the SAR domain, prior UAV-based studies have investigated target search, localization, and sensing-path design in disaster environments~\cite{Perazzo16:Drone,Yuan22:A,Ebrahimi20:Autonomous,Liu12:HAWK}. Perazzo et al.~\cite{Perazzo16:Drone} proposed the LocalizerBee framework, which discretizes the search area using triangular grids determined by the UAV communication range. Yuan et al.~\cite{Yuan22:A} considered NLOS localization using a rectangular-grid representation and a trajectory designed to provide full spatial coverage. Liu et al.~\cite{Liu12:HAWK} employed Moore space-filling curves for deterministic search-area traversal. These approaches provide structured sensing or routing strategies, but do not directly optimize the selection of a minimum set of UAV sensing waypoints.

Beyond SAR-specific applications, coverage path planning (CPP) studies have addressed multi-region and energy-aware coverage planning~\cite{Xie20:Path,Li20:A}, while evolutionary approaches such as genetic algorithms have also been explored~\cite{Yuan22:Global}. Despite this progress, these methods generally focus on complete coverage routes or sweep patterns over predefined locations. Our formulation instead considers an extended $k$-coverage problem in which each target grid point must be covered by at least $k$ distinct UAV waypoints and the waypoint subset itself is jointly optimized.

The multiple-coverage requirement underlying our formulation is closely related to the classical $k$-coverage problem, which has been studied primarily in the context of wireless sensor networks (WSNs)~\cite{Hefeeda07:Randomized,Ahmadi14:An,Yu17:Coverage}. In conventional WSN $k$-coverage, the objective is to ensure that every point in a target region is covered by at least $k$ distinct sensors while minimizing the required sensing resources. A representative formulation is given as follows:
\begin{problem}[$k$-Coverage Problem in WSN \cite{Hefeeda07:Randomized}]
Given a target area, a desired coverage degree \(k \geq 1\), and a set of sensors with fixed sensing ranges, determine the minimum number (and placement) of sensors such that every point in the area lies within the sensing range of at least \(k\) distinct sensors.
\end{problem}

\begin{figure}
    \centering
    \includegraphics[width=0.55\linewidth]{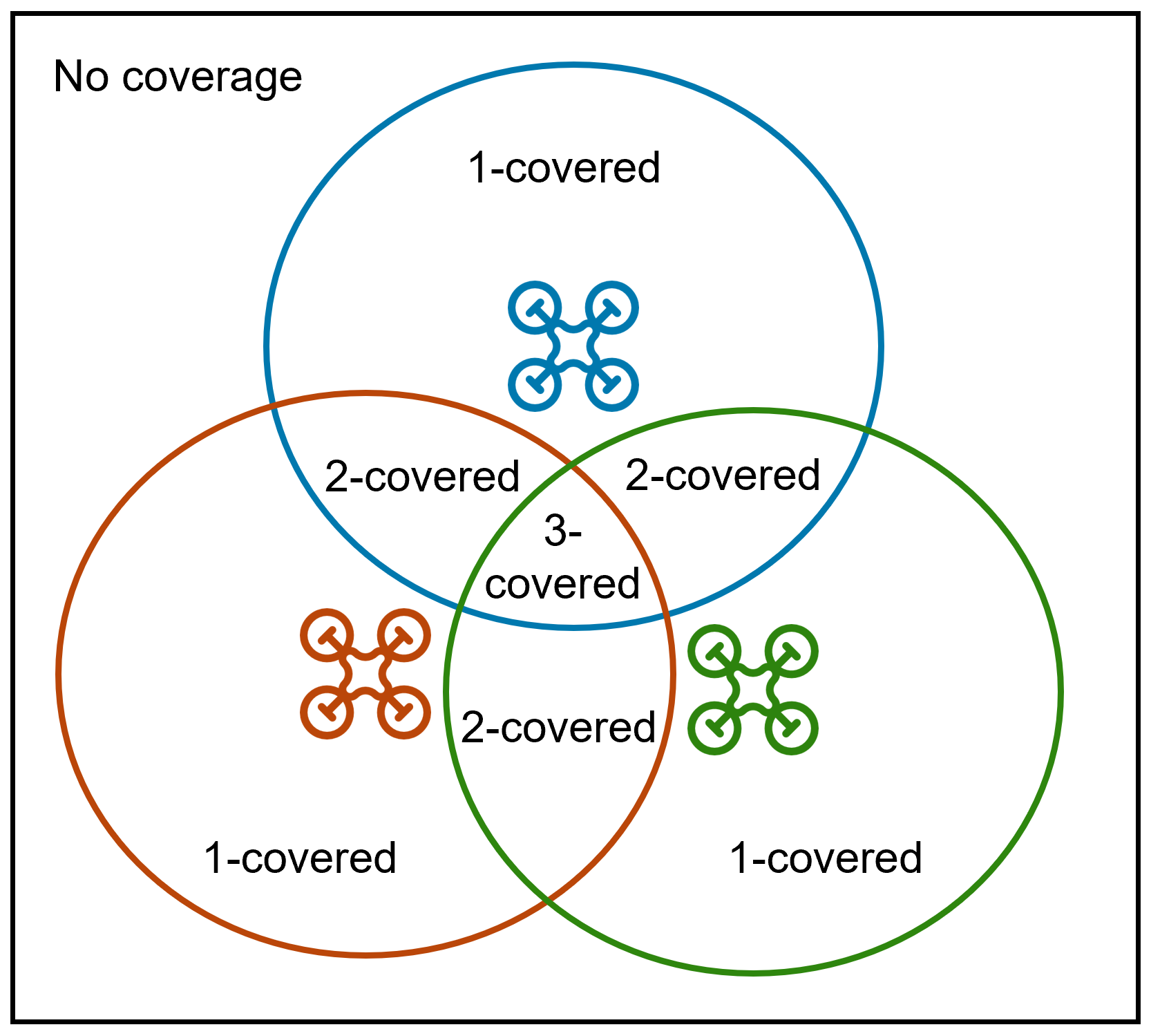}
    \caption{Illustration of $k$-level coverage.}
    \label{fig:CoverageLevel}
\end{figure}

Fig.~\ref{fig:CoverageLevel} illustrates an example of $k$-coverage involving three waypoints, where regions covered by one, two, or all three sensing nodes are labeled accordingly. While the classical $k$-coverage problem has been well studied in WSN applications, directly applying the conventional WSN $k$-coverage model to UAV-based SAR scenarios is not straightforward due to a fundamental difference in spatial configuration. In WSN formulations, sensors are typically deployed within the target region, whereas in SAR missions, UAVs operate only within flight-accessible free space (e.g., above or around buildings), while missing persons may reside inside structures. Consequently, the practical environment can be naturally decomposed into (i) a UAV-navigable region and (ii) a target region where coverage must be guaranteed. To faithfully capture this distinction, we extend the classical $k$-coverage formulation to explicitly model these two regions and to allow coverage to be provided only from navigable waypoints. This extended $k$-coverage model serves as the foundation of our waypoint optimization framework presented in Section~\ref{Sec:Method}.

\subsection{Quantum optimization}
\label{Sec:QuantumOptimization}
QAOA is a hybrid variational algorithm that alternates applications of unitaries generated by a problem\mbox{-}specific cost Hamiltonian $H_C$ and a mixer Hamiltonian $H_M$.
A classical optimizer then updates the circuit parameters to minimize the expectation value of $H_C$ \cite{Farhi2014:QAOA}.
For combinatorial optimization, one typically encodes the problem as a QUBO model with binary variables $x_i \in \{ 0, 1 \}$ and maps it to an equivalent Ising-form cost Hamiltonian with spin variables $z_i \in \{ \pm 1 \}$.
Here, \(H_C\) encodes the objective function to be minimized, while \(H_M\) induces transitions among computational-basis states to efficiently explore the search space.

Concretely, let a QUBO objective function be
\begin{equation}
    f(\mathbf{x}) \;=\; \mathbf{x}^\top Q\,\mathbf{x} \;+\; \mathbf{c}^\top \mathbf{x} \;+\; c_0,
\qquad \mathbf{x}\in\{0,1\}^n,\quad Q=Q^\top ,
\end{equation}
where $Q \in \mathbb{R}^{n \times n}$ is the matrix of quadratic coefficients, $\mathbf{c} \in \mathbb{R}^n$ is a vector of linear coefficients, and $c_0 \in \mathbb{R}$ is a constant offset.
Using the standard affine transformation in terms of the Pauli\mbox{-}$Z$ operator,
\begin{equation}
    x_i \mapsto \tfrac{1}{2}\!\left(I - Z_i\right), 
    \qquad
    x_i x_j \mapsto \tfrac{1}{4}\!\left(I - Z_i - Z_j + Z_i Z_j\right),
\end{equation}
we obtain a diagonal Ising\mbox{-}form cost Hamiltonian
\begin{equation}
    H_C = \sum_{i<j} J_{ij}\, Z_i Z_j \;+\; \sum_i h_i\, Z_i \;+\; E_0 I .
\end{equation}
The coefficients $J_{ij}$, $h_i$, and $E_0$ are obtained
directly from $Q$, $\mathbf{c}$, and $c_0$ under the mapping above.

Given a cost Hamiltonian $H_C$ and a mixer Hamiltonian $H_M$,
we define the parameterized unitaries
$U_C(\gamma)=e^{-i\gamma H_C}$ and
$U_M(\beta)=e^{-i\beta H_M}$.
In this work, we use the standard $X$-mixer,
$H_M=\sum_{i=1}^{n}X_i$.
A depth-$p$ QAOA ansatz is constructed by alternating these
operators $p$ times with layer-specific parameters
$\{(\gamma_k,\beta_k)\}_{k=1}^{p}$:
\begin{equation}
|\psi_p(\boldsymbol{\gamma},\boldsymbol{\beta})\rangle
=
U_M(\beta_p)U_C(\gamma_p)\cdots
U_M(\beta_1)U_C(\gamma_1)|+\rangle^{\otimes n},
\end{equation}
where
$\boldsymbol{\gamma}=(\gamma_1,\ldots,\gamma_p)$,
$\boldsymbol{\beta}=(\beta_1,\ldots,\beta_p)$, and
$\ket{+}^{\otimes n}
=
H^{\otimes n}\ket{0}^{\otimes n}$.
The classical optimizer updates
$\boldsymbol{\gamma}$ and $\boldsymbol{\beta}$ to minimize
the expected energy
\begin{equation}
\label{eq:qaoa-energy}
E(\boldsymbol{\gamma},\boldsymbol{\beta})
=
\big\langle
\psi_p(\boldsymbol{\gamma},\boldsymbol{\beta})
\big|
H_C
\big|
\psi_p(\boldsymbol{\gamma},\boldsymbol{\beta})
\big\rangle,
\end{equation}
which is estimated from repeated circuit executions.
At the final parameter vector, the circuit is sampled $m$
times in the computational basis, yielding QUBO assignments
$\{\mathbf{x}_s\}_{s=1}^{m}$ with
$\mathbf{x}_s\in\{0,1\}^{n}$.
For the application-level results, we define the selected
solution $\widehat{\mathbf{x}}$ using the best-of-shots
(BoS) rule:
\begin{equation}
\label{eq:bos}
\widehat{\mathbf{x}}
\in
\arg\min_{\mathbf{x}\in
\{\mathbf{x}_1,\ldots,\mathbf{x}_m\}}
f(\mathbf{x}).
\end{equation}
If a global QUBO minimizer, equivalently a computational-basis ground-state assignment of $H_C$, is sampled, the BoS rule retains one such minimizer; otherwise, it returns the lowest-cost sampled assignment.

QAOA has been investigated for diverse combinatorial optimization problems, including MaxCut \cite{Farhi2014:QAOA}, knapsack \cite{Kea23:Leveraging}, dominating set \cite{Li25:Quantum}, and the minimum vertex cover problem (MVCP) \cite{Zhang22:Applying,Li25:Performance}.
For MVCP, Zhang et al.~\cite{Zhang22:Applying} reported optimal or near-optimal solutions on small simulated graphs, while Li et al.~\cite{Li25:Performance} extended the approach to the weighted case using QAOA+.
Application-oriented QAOA studies have further addressed 6G routing \cite{Bouchmal25:Quantum}, vehicle routing \cite{Azad22:Solving}, job-shop scheduling \cite{Kurowski23:Application}, and electric-vehicle charging \cite{Kea23:Leveraging}.
Bouchmal et al.~\cite{Bouchmal25:Quantum} derived an asymptotic improvement for multi-objective routing; Azad et al.~\cite{Azad22:Solving} reported near-optimal routes on small instances without demonstrating a runtime advantage; Kurowski et al.~\cite{Kurowski23:Application} obtained optimal schedules for toy benchmarks and observed depth-dependent parameter patterns; and Kea et al.~\cite{Kea23:Leveraging} reported strong noise-free performance with degradation on NISQ hardware.
Among these studies, QAOA-based MVCP is most closely related to our formulation; however, it is limited to single-coverage graph problems and classical simulation, whereas localization requires minimum $k$-coverage.

\begin{table*}[t]
\centering
\caption{Comparison of the proposed framework with representative related works. The proposed method is distinguished by extended 3-coverage waypoint-subset optimization, explicit separation of the UAV-navigable region $\mathcal{U}$ and target region $\mathcal{T}$, training-free deployment, and validation on real quantum hardware.}
\label{tab:related_work_comparison}
\renewcommand{\arraystretch}{1.6}
\setlength{\tabcolsep}{5pt}
\setlength{\aboverulesep}{0pt}
\setlength{\belowrulesep}{0pt}
\begin{tabular}{
  >{\raggedright\arraybackslash}m{2.4cm}   
  >{\centering\arraybackslash}m{2.5cm}    
  >{\centering\arraybackslash}m{1.5cm}    
  >{\centering\arraybackslash}m{1.8cm}    
  >{\centering\arraybackslash}m{1.5cm}    
  >{\centering\arraybackslash}m{1.5cm}    
  >{\centering\arraybackslash}m{3.2cm}    
}
\toprule
\noalign{\vspace{4pt}}
\rule{0pt}{4ex}\textbf{Work}
  & \rule{0pt}{4ex}\textbf{Application}
  & \makecell{\textbf{Waypoint} \\ \textbf{Subset} \\ \textbf{Optimized?}}
  & \makecell{\textbf{Explicit} \\ \textbf{Min.} \\ \textbf{$k$-Coverage?}}
  & \rule{0pt}{4.5ex}\makecell{\textbf{Separate} \\ $\mathcal{U}$ \textbf{\&} $\mathcal{T}$\textbf{?}}
  & \rule{0pt}{4.5ex}\makecell{\textbf{Training-} \\ \textbf{Free?}}
  & \rule{0pt}{4ex}\textbf{Solver / Validation} \\
\noalign{\vspace{3pt}}
\midrule
\noalign{\vspace{1pt}}
Perazzo et al.~\cite{Perazzo16:Drone}
  & UAV-based search \& localization
  & \texttimes & \texttimes & \texttimes & \checkmark
  & Deterministic grid construction \\
Yuan et al.~\cite{Yuan22:A}
  & UAV-based localization in NLOS
  & \texttimes & \texttimes & \texttimes & \checkmark
  & Deterministic grid construction \\
Ebrahimi et al.~\cite{Ebrahimi20:Autonomous}
  & UAV-based localization
  & \texttimes & \texttimes & \texttimes & \texttimes
  & Reinforcement learning \\
CPP studies~\cite{Xie20:Path, Vasquez18:Coverage, Li20:A, Yuan22:Global, Kapanoglu12:Pattern, Ab09:Wireless, Chen22:Coverage}
  & UAV area coverage
  & \texttimes & \texttimes & --- & \checkmark
  & Classical heuristic \\
WSN $k$-coverage studies~\cite{Hefeeda07:Randomized, Ahmadi14:An, Yu17:Coverage, Li13:A, Elhoseny18:Optimizing}
  & Wireless sensor network coverage
  & --- & \checkmark & \texttimes & \checkmark
  & Classical heuristic \\
QAOA-based MVCP~\cite{Zhang22:Applying, Li25:Performance}
  & Combinatorial graph optimization
  & --- & \texttimes & --- & \checkmark
  & QAOA / Classical simulation \\
\noalign{\vspace{2pt}}
\midrule
\textbf{This work}
  & \textbf{RF-based UAV SAR}
  & $\bm{\checkmark}$ & $\bm{\checkmark}$ & $\bm{\checkmark}$ & $\bm{\checkmark}$
  & \textbf{QAOA / Simulator + real quantum HW} \\
 
\bottomrule
\multicolumn{7}{l}{%
  \footnotesize
  \checkmark: satisfied \quad
  \texttimes: not satisfied \quad
  ---: not applicable
}
\end{tabular}
\end{table*}

In this work, we address the multiple-coverage conditions used to support RF-based localization.
We formulate the extended $k$-coverage waypoint optimization problem, fundamental to SAR, as a QUBO instance, solve it with QAOA, and experimentally validate the approach on both a noise-free quantum circuit simulator and IBM’s 127-qubit Eagle processor. Table~\ref{tab:related_work_comparison} summarizes the main differences between the proposed framework and representative prior studies.

\section{Quantum Algorithm for UAV Waypoint Optimization} \label{Sec:Method}

\begin{figure*}[tbp]
    \centering
    \includegraphics[width=0.8\textwidth]{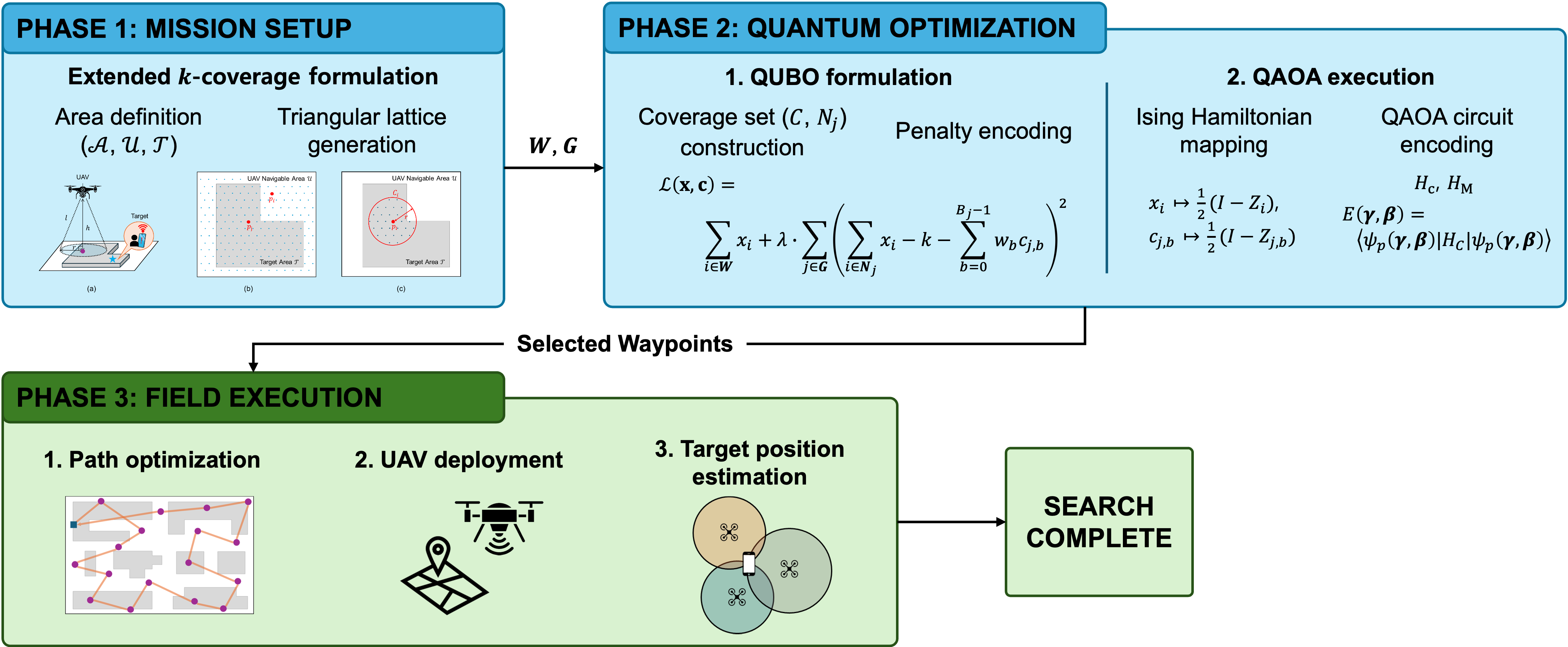}
\caption{End-to-end operational workflow of the proposed framework. The system proceeds through three sequential phases: (1) mission setup, where the search area is discretized into a triangular lattice to yield $\boldsymbol{W}$ and $\boldsymbol{G}$; (2) quantum optimization, where the extended $k$-coverage problem is formulated as a QUBO model and solved via QAOA to determine the selected waypoint subset; and (3) field execution, where the optimized waypoints are utilized for UAV path planning and RF-based target localization.}
    \label{fig:operational_workflow}
\end{figure*}

This section presents the proposed QAOA-based framework for UAV waypoint optimization. Fig.~\ref{fig:operational_workflow} illustrates the end-to-end workflow comprising three phases: mission setup, quantum optimization, and field execution. The following subsections detail the algorithmic components of the first two phases. Specifically, we formulate the waypoint-selection task as an extended $k$-coverage problem, derive its QUBO and corresponding QAOA-compatible Hamiltonian, and solve the resulting formulation using the standard QAOA framework on near-term quantum devices.

\subsection{Extended $k$-coverage problem} \label{Sec:kCoverageProb}

We formulate the UAV waypoint optimization task as an extended version of the $k$-coverage problem. Under the assumed 2D localization setting, the target position contributes two unknown spatial coordinates, while TOA- or RSS-based localization may involve one additional modality-dependent scalar parameter, such as a clock offset or transmit power. This parameterization motivates the use of at least three spatially distinct measurements, and we therefore set $k=3$ as a minimum measurement-multiplicity requirement. Practical localization performance additionally depends on sensing geometry, measurement noise, and model validity. To formalize the extended $k$-coverage problem, we define:

\begin{itemize}
    \item Let \(\mathcal{A}\) denote the entire two-dimensional search area.
    \item Let \(\mathcal{U} \subseteq \mathcal{A}\) represent the ground projection of the UAV-navigable area.
    \item Let \(\mathcal{T} \subseteq \mathcal{A}\) represent the target area in which missing persons may exist.
\end{itemize}

\begin{figure*}[t]
    \centering
    \includegraphics[width=0.85\textwidth]{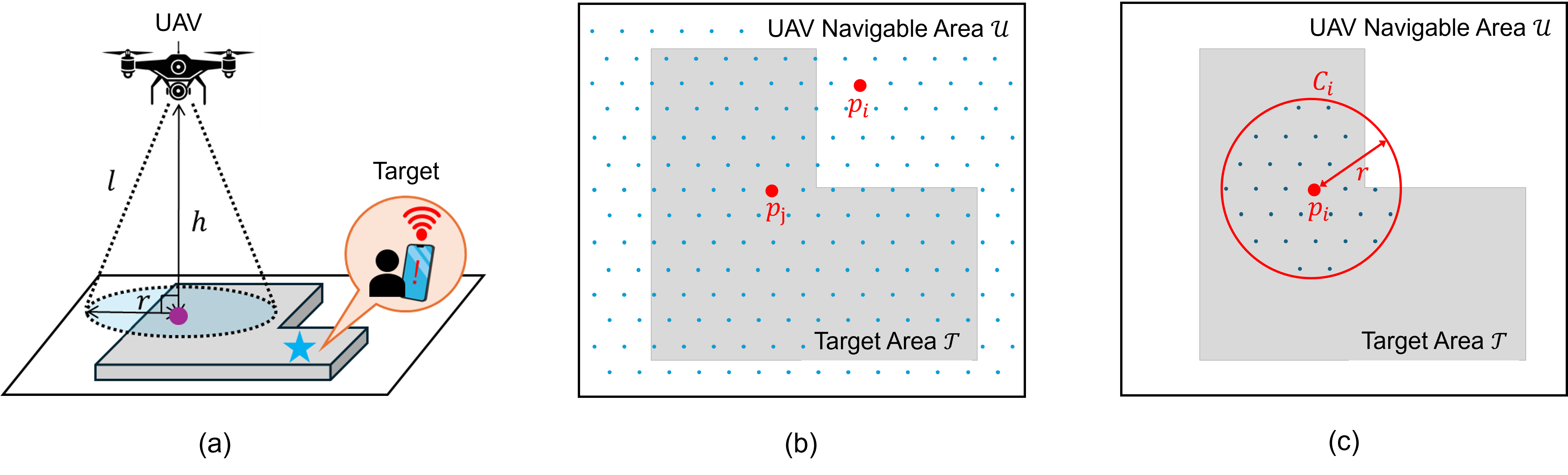}
    \caption{Illustration of the extended $k$-coverage waypoint optimization problem.  
(a) UAVs collect RF measurements emitted by targets (e.g., smartphones) from spatially distinct waypoints for target localization. A purple dot denotes the ground projection of a candidate UAV waypoint, and a cyan star represents a candidate target location.
(b) The search area is discretized into grid points. Cyan points denote candidate UAV waypoints in the UAV-navigable area $\mathcal{U}$, and the shaded region represents the target area $\mathcal{T}$. For example, $p_i$ is a candidate waypoint and $p_j$ is a target point.
(c) For each waypoint $p_i \in \boldsymbol{W}$, its coverage set $C_i$ is defined as the set of target points within radius $r$, i.e., $C_i = \{p_j \in \boldsymbol{G} \mid d(p_i, p_j) \leq r\}$.}
    \label{fig:ProblemDescription}
\end{figure*}

The search area $\mathcal{A}$, navigable region $\mathcal{U}$, and target region $\mathcal{T}$ are assumed to be available at mission time from map sources such as satellite imagery or public geographic databases. While UAVs operate at altitude, their coverage can be analyzed in terms of a projected ground area. In other words, although UAVs fly in three-dimensional (3D) space, the effective region within which they can receive RF signals from ground-level targets can be represented on the two-dimensional (2D) ground plane. This relationship follows from geometric projection, where the UAV's ground coverage radius is given by \( r = \sqrt{l^2 - h^2} \), based on its communication range \( l \) and altitude \( h \), as shown in Fig.~\ref{fig:ProblemDescription}(a).

To discretize the entire area $\mathcal{A}$ into a set of grid points, we generate a triangular lattice with grid spacing \(\delta\). The lattice is defined by the basis vectors
\[
\mathbf{e}_1 = \left(\delta, 0\right), \qquad
\mathbf{e}_2 = \left(\frac{\delta}{2},\, \frac{\sqrt{3}}{2}\delta\right),
\]
and lattice points are generated as $p_{u,v} = u\mathbf{e}_1 + v\mathbf{e}_2$, for integer indices \(u, v\).

Let $P$ denote the total number of lattice points generated within $\mathcal{A}$. All such lattice points are collected and assigned sequential indices $i \in \{1, \ldots, P\}$, yielding the set
\[
\boldsymbol{p} = \{ p_i \}_{i=1}^{P}.
\]
Thus, each point $p_i$ is uniquely represented by its index $i$ for the optimization formulation, and by its axial coordinate pair $(u, v)$ for its geometric location on the lattice.

Each grid point may serve as (i) a candidate UAV waypoint if it lies in \(\mathcal{U}\), 
and/or (ii) a potential target location if it lies in \(\mathcal{T}\). 
Accordingly, we define
\begin{itemize}
    \item \(\boldsymbol{W} \subseteq \boldsymbol{p}\): the set of candidate UAV waypoints (i.e., grid points in \(\mathcal{U}\)),
    \item \(\boldsymbol{G} \subseteq \boldsymbol{p}\): the set of potential target locations (i.e., grid points in \(\mathcal{T}\)).
\end{itemize}

Note that some grid points may belong to both \(\boldsymbol{W}\) and \(\boldsymbol{G}\). Fig.~\ref{fig:ProblemDescription}(b) depicts this setup as a top-down view of the 3D environment shown in Fig.~\ref{fig:ProblemDescription}(a). For example, if a target is located somewhere inside the gray building region, then \(\boldsymbol{G}\) includes the grid points inside the building (i.e., the sky-blue grids within the gray area), while \(\boldsymbol{W}\) includes the UAV-accessible points located both above and around the building (i.e., all sky-blue grids within the rectangular boundary).

We now formally define the waypoint selection problem:

\begin{problem}[Extended \(k\)-Coverage Problem]
\label{Prob:ExtendedKCoverage}
Given a set of waypoint grid points \(\boldsymbol{W}\), a set of target grid points \(\boldsymbol{G}\), and a radius \(r\), determine a minimal subset \(\boldsymbol{O} \subseteq \boldsymbol{W}\) such that every \(p_j \in \boldsymbol{G}\) is within distance \(r\) of at least \(k\) distinct waypoints in \(\boldsymbol{O}\).
\end{problem}

Note that our formulation generalizes the conventional $k$-coverage problem. In particular, when $\mathcal{U} = \mathcal{T}$, Problem~\ref{Prob:ExtendedKCoverage} reduces exactly to the conventional $k$-coverage problem. Since the conventional $k$-coverage problem is NP-hard, the extended $k$-coverage problem is also NP-hard. A formal proof is provided in Appendix~A of the supplementary material.

\subsection{QUBO formulation}
\label{subsec:QUBO_formulation}
To make the problem amenable to quantum optimization, we reformulate it as a QUBO model. This process involves three main steps: 1) defining the objective function by introducing a binary decision vector,  
2) formalizing the coverage constraint via set-based representation, and 3) converting the constrained formulation into a QUBO model.

\subsubsection{Objective Function}
The objective of the extended $k$-coverage problem is to find a minimum subset of candidate UAV waypoints $\boldsymbol{W}$. We first define a binary decision vector $\mathbf{x} \in \{0,1\}^{|\boldsymbol{W}|}$, where $x_i = 1$ indicates that waypoint $p_i \in \boldsymbol{W}$ is selected, and $x_i = 0$ otherwise. The goal is to minimize the total number of selected waypoints, which is given by the sum of the entries in $\mathbf{x}$, as shown later in Eq.~(\ref{eq:ConstrainedOptimization}).

\subsubsection{Coverage Constraint}
Next, we enforce the coverage constraint such that every target point in $\boldsymbol{G}$ must be covered by at least $k$ selected waypoints. To do so, we define a collection of coverage sets $\mathcal{C} = \{\boldsymbol{C}_i \mid p_i \in \boldsymbol{W}\}$, where each set $\boldsymbol{C}_i \subseteq \boldsymbol{G}$ represents the target grid points covered by UAV waypoint $p_i$:
\begin{equation}
\boldsymbol{C}_i = \{p_j \in \boldsymbol{G} \mid d(p_i, p_j) \leq r\},\label{eq:setC}
\end{equation}
with $d(\cdot, \cdot)$ denoting the Euclidean distance between two grid points. Fig.~\ref{fig:ProblemDescription}(c) illustrates Eq.~(\ref{eq:setC}): for each candidate waypoint $p_i$, a circle of radius $r$ is drawn, and $\boldsymbol{C}_i$ consists of the target grid points located within the circle and inside the target area.

To represent the coverage requirement, we define, for each target point $p_j \in \boldsymbol{G}$, the index set:
\begin{equation}
\boldsymbol{N}_j = \{i \mid p_j \in \boldsymbol{C}_i\},
\end{equation}
which indicates the indices of waypoints that can cover $p_j$.
We assume that $|\boldsymbol{N}_j|\geq k$ for every
$j\in\boldsymbol{G}$; otherwise, Problem ~\ref{Prob:ExtendedKCoverage} is infeasible
and the instance is excluded before QUBO construction.
The resulting constrained optimization problem is formulated as:
\begin{equation}
\begin{aligned}
\min_{\mathbf{x}} \quad & \sum_{i \in \boldsymbol{W}} x_i  \\
\text{s.t.} \quad & \sum_{i \in \boldsymbol{N}_j} x_i \geq k, \quad \forall j: p_j \in \boldsymbol{G}.
\end{aligned}\label{eq:ConstrainedOptimization}
\end{equation} 

\subsubsection{QUBO Model}
\label{subsubsec:QUBO_Model}
To cast Eq.~(\ref{eq:ConstrainedOptimization}) into a QUBO model, we transform the inequality constraints into equality constraints by introducing nonnegative slack variables \(s_j\) that capture the surplus in coverage:
\begin{equation} 
\label{eqn:InequalityToEquality}
\sum_{i \in \boldsymbol{N}_j} x_i - s_j = k, \quad s_j \geq 0.
\end{equation}

This implies:
\begin{equation}
\label{eqn:slackVariable}
s_j = \max \left\{ 0, \sum_{i \in \boldsymbol{N}_j} x_i - k \right\}.
\end{equation}
Each slack variable \(s_j\) is encoded into a binary representation to conform to the QUBO formulation. Suppose the maximum value that \(s_j\) can attain is \(n_j - k\), where \(n_j = |\boldsymbol{N}_j|\).
Then we can express \(s_j\) as a weighted sum of binary variables:
\begin{equation}
\label{eqn:slackBinary}
s_j = \sum_{b=0}^{B_j-1} w_b\, c_{j,b}, \quad c_{j,b} \in \{0,1\}, \quad w_b=2^b
\end{equation}
where  $B_j = \lceil \log_2(n_j - k + 1) \rceil$ is the minimum number of binary variables required to represent the range of \(s_j\).
By jointly optimizing the slack variables with the binary decision vector $\mathbf{x}$, this formulation penalizes coverage violations in the QUBO objective, effectively discouraging infeasible solutions and promoting feasible configurations.

Substituting this into the objective yields the QUBO formulation:
\begin{equation}
\label{eqn:QUBO_final}
\mathcal{L}(\mathbf{x}, \mathbf{c}) = \sum_{i\in \boldsymbol{W}} x_i  + \lambda \cdot \sum_{j\in \boldsymbol{G}} \left( \sum_{i \in \boldsymbol{N}_j} x_i - k - \sum_{b=0}^{B_j-1} w_b c_{j,b} \right)^2,
\end{equation}
where $\lambda>0$ is a penalty coefficient that penalizes violations of the $k$-coverage requirement.
For the uniform-penalty formulation in Eq.~(\ref{eqn:QUBO_final}), $\lambda>|\boldsymbol{W}|$ is a sufficient condition for exactness: the decision component of every global minimizer is feasible for Problem~\ref{Prob:ExtendedKCoverage} and minimizes the waypoint cardinality over the feasible set.
A formal proof is provided in Proposition~2 in Appendix~B of the supplementary material.

Eq.~(\ref{eqn:QUBO_final}) presents the canonical uniform-penalty QUBO used for the exactness analysis.
For the simulator and hardware experiments, however, the QUBO instances were generated using \texttt{QuadraticProgramToQubo} in Qiskit Optimization 0.7.0 with automatic penalty selection, which may assign different penalty coefficients at different constraint-conversion stages.
Accordingly, the theoretical condition on the common penalty coefficient $\lambda$ applies to the canonical formulation in Eq.~(\ref{eqn:QUBO_final}); in the controlled penalty-sensitivity study in Section~\ref{sec:Simulator_analysis}, we explicitly use a common scalar $\lambda$ across the conversion stages to evaluate this condition.

\subsection{QAOA‑based optimization}
In this section, we convert the waypoint‑optimization QUBO in Eq.~\eqref{eqn:QUBO_final} into an Ising‑form cost Hamiltonian. 
Each binary variable is mapped via the standard affine mapping in terms of the Pauli‑$Z$ operator:
\begin{equation}
    \label{eq:op-mapping}
    x_i \;\mapsto\; \tfrac{1}{2}\!\left(I - Z_i\right),\qquad
    c_{j,b} \;\mapsto\; \tfrac{1}{2}\!\left(I - Z_{j,b}\right).
\end{equation}
Under this mapping, the cardinality term in \eqref{eqn:QUBO_final}, $\sum_{i\in \boldsymbol{W}} x_i$, transforms as
\begin{equation}
    \sum_{i\in \boldsymbol{W}} x_i \;\mapsto\; 
    \tfrac{|\boldsymbol{W}|}{2}\,I \;-\; \tfrac{1}{2}\sum_{i\in \boldsymbol{W}} Z_i.
\end{equation}
Next, we apply Eq.~\eqref{eq:op-mapping} term by term to the coverage-constraint expression in Eq.~\eqref{eqn:QUBO_final}, $\sum_{i \in \boldsymbol{N}_j} x_i \;-\; k \;-\; \sum_{b=0}^{B_j-1} w_b\, c_{j,b}$, obtaining
\begin{equation}
    \begin{aligned}
    \sum_{i \in \boldsymbol{N}_j} x_i
    &\;\mapsto\;
    \sum_{i \in \boldsymbol{N}_j} \tfrac{1}{2}\!\left(I - Z_i\right)
    \;=\; \tfrac{|\boldsymbol{N}_j|}{2}\, I \;-\; \tfrac{1}{2}\!\sum_{i \in \boldsymbol{N}_j} Z_i, \\[4pt] 
    k
    &\;\mapsto\; k\,I, \\[4pt]
    \sum_{b=0}^{B_j-1} w_b\, c_{j,b}
    &\;\mapsto\;
    \sum_{b=0}^{B_j-1} w_b\, \tfrac{1}{2}\!\left(I - Z_{j,b}\right)\\
    &=\; \tfrac{1}{2}\!\sum_{b=0}^{B_j-1} w_b\, I \;-\; \tfrac{1}{2}\!\sum_{b=0}^{B_j-1} w_b\, Z_{j,b}.
    \end{aligned}
\end{equation}
Collecting identity and $Z$-operator terms, the coverage-constraint operator becomes
\begin{equation}
\begin{aligned}
&\sum_{i\in \boldsymbol{N}_j} x_i - k - \sum_{b=0}^{B_j-1} w_b\, c_{j,b} \\
&\mapsto \left(\frac{|\boldsymbol{N}_j|}{2} - \frac{1}{2}\sum_{b=0}^{B_j-1} w_b - k\right) I
      - \frac{1}{2}\sum_{i\in \boldsymbol{N}_j} Z_i \\
      &\qquad \qquad \qquad + \frac{1}{2}\sum_{b=0}^{B_j-1} w_b\, Z_{j,b}.
\end{aligned}
\end{equation}
By squaring and summing these coverage-constraint terms over \(j\in \boldsymbol{G}\), we obtain the operator form of the quadratic penalty in Eq.~\eqref{eqn:QUBO_final}.
Adding the mapped cardinality term yields the Ising‑form cost Hamiltonian $H_C$:
\begin{equation}
\label{eq:HC_final}
\begin{aligned}
H_C
&= \frac{|\boldsymbol{W}|}{2}\, I \;-\; \frac{1}{2}\sum_{i\in \boldsymbol{W}} Z_i \\
&\quad + \lambda \sum_{j\in \boldsymbol{G}}
\Biggl[
\left(\frac{|\boldsymbol{N}_j| - \sum_{b=0}^{B_j-1} w_b}{2} - k\right) I \\
&\quad - \frac{1}{2}\sum_{i\in \boldsymbol{N}_j} Z_i
+ \frac{1}{2}\sum_{b=0}^{B_j-1} w_b\, Z_{j,b}
\Biggl]^2.
\end{aligned}
\end{equation}
Identity‑proportional terms act as global energy offsets and are dropped without loss.
Let \(n =|\boldsymbol{W}| + \sum_{j\in\boldsymbol{G}} B_j\) denote the total number of qubits. 
The standard \(X\)-mixer acting on all \(n\) qubits is
\[
H_M \;=\; \sum_{q=1}^{n} X_q .
\]
With the cost Hamiltonian $H_C$ defined above, the depth-\(p\) QAOA ansatz is
\begin{equation}
\label{eqn:QAOAAnsatz}
\ket{\psi_p(\boldsymbol{\gamma},\boldsymbol{\beta})}
= \Bigg[\prod_{k=1}^{p} e^{-i \beta_k H_M}\, e^{-i \gamma_k H_C}\Bigg]
\ket{+}^{\otimes n},
\end{equation}
where \(\boldsymbol{\gamma}=(\gamma_1,\ldots,\gamma_p)\) and \(\boldsymbol{\beta}=(\beta_1,\ldots,\beta_p)\).
The parameters are chosen to minimize
\begin{equation}
E(\boldsymbol{\gamma},\boldsymbol{\beta})
= \bra{\psi_p(\boldsymbol{\gamma},\boldsymbol{\beta})} H_C \ket{\psi_p(\boldsymbol{\gamma},\boldsymbol{\beta})},
\end{equation}
thereby steering the quantum state toward the ground state of $H_C$.

\section{Simulator-Based Feasibility and Optimality Validation}
\label{sec:Simulator_analysis}

Due to the limited fidelity and high error rates of current NISQ hardware, we first evaluated the proposed framework using a noise-free state-vector simulator. This controlled environment isolates the algorithmic behavior from hardware-induced errors. The simulator platform comprised two NVIDIA RTX 6000 GPUs (48 GB), one NVIDIA L40S GPU, and one NVIDIA A100 GPU (48 GB), supporting state-vector simulations of circuits containing up to 34 qubits under the available GPU-memory constraint.
All simulations were implemented using the Qiskit Aer \texttt{Sampler} primitive.
Unless otherwise stated, the baseline configuration used a QAOA depth of $p=2$ and 1024 shots per objective-function evaluation.
The classical optimization stage employed the constrained
optimization by linear approximations (COBYLA) algorithm with an objective-function evaluation budget of one per run.
The QUBO instances were constructed using the conversion procedure described in Section~\ref{Sec:Method}.

All simulator-based experiments were restricted to at most 34 qubits by the available GPU memory required for state-vector simulation.
Increasing the size of the target area or using a finer grid discretization directly increases the number of grid points and coverage interactions, which in turn enlarges the number of binary decision and slack variables and quickly exceeds the available qubit budget.
Within this limitation, we therefore aimed to evaluate the proposed formulation across as diverse a set of target-area geometries and spatial configurations as possible.
A detailed analysis of quantum resource requirements and scalability is provided in Section~\ref{sec:QuantumResourceRequirements}.

\begin{figure*}[t]
    \centering
    \includegraphics[width=0.85\textwidth]{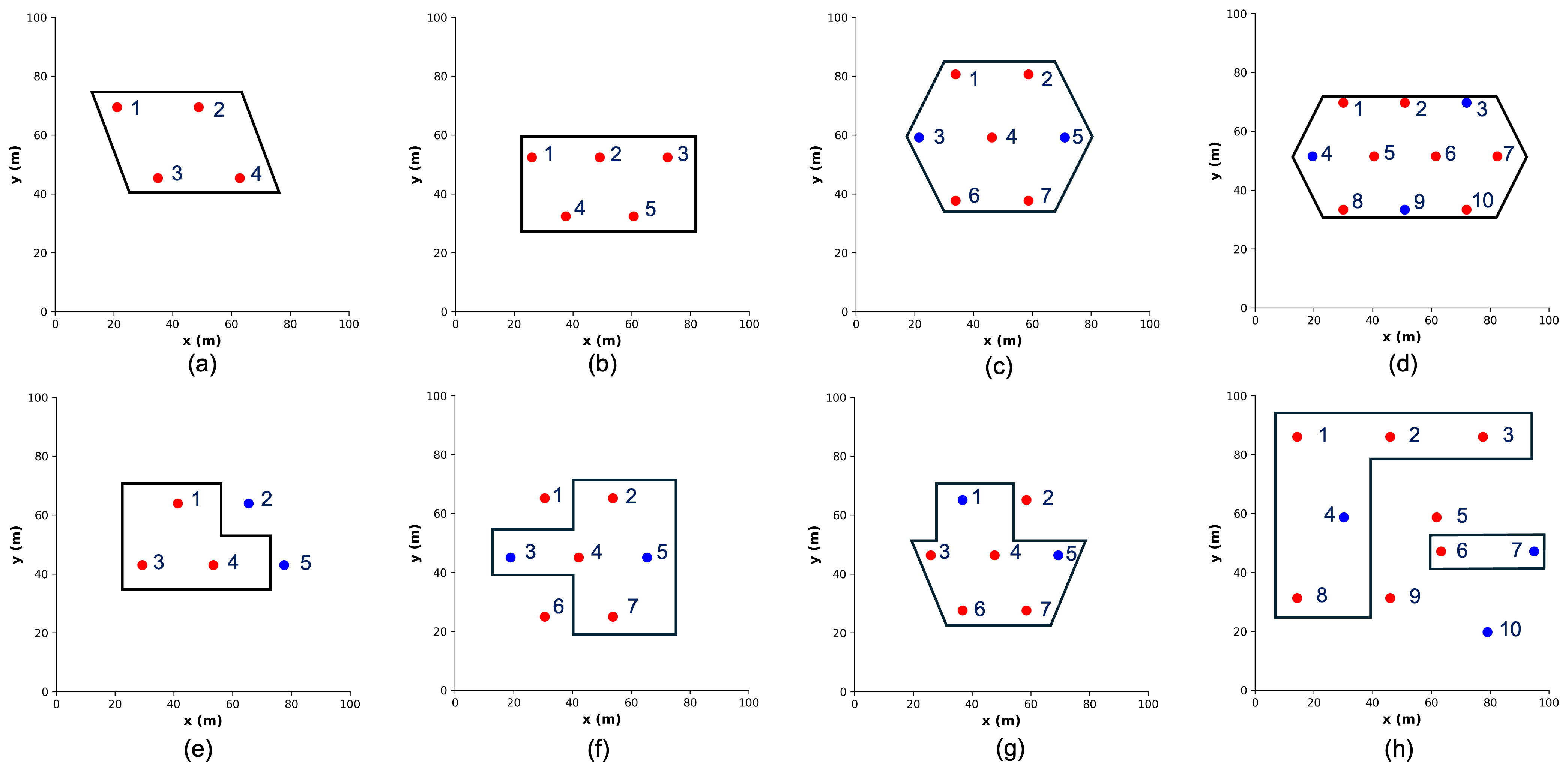}
    \caption{Eight test scenarios used in simulator-based evaluation. Blue and red points denote candidate UAV waypoints, black boundaries represent potential target areas, and red points mark the waypoint subset selected by QAOA.}
    \label{fig:EmulatorScenarios}
\end{figure*}

We constructed eight test scenarios to reflect a range of target-area geometries and spatial configurations, as illustrated in Fig.~\ref{fig:EmulatorScenarios}.
Blue and red points indicate candidate waypoints, with the red points denoting the subset selected by QAOA; black boundaries represent potential target areas.
These scenarios vary in convexity, grid density, and the spatial distribution of UAV waypoints---specifically, whether candidate waypoints are located exclusively within or also beyond the boundaries of the target region.
The UAV ground-projected communication radius was fixed at $r = 35$~m for all scenarios, which was found to consistently enable feasible 3-coverage across all test cases while remaining within the simulator’s qubit constraint.

\begin{table*}[htbp]
\caption{Simulator evaluation results obtained with the Qiskit Aer simulator for all eight test scenarios. For each scenario (a–h), the grid spacing ($\delta$), the number of candidate waypoints ($|\boldsymbol{W}|$), the selected UAV waypoints returned by QAOA, and the known minimum number of waypoints required for 3-coverage are reported. In all cases, the QAOA-based solution satisfies both feasibility (i.e., ensuring 3-coverage of all targets) and optimality (i.e., selecting the minimal number of waypoints).}
\centering
\begin{tabular}{lcccccccc}
\toprule
 & \textbf{a} & \textbf{b} & \textbf{c} & \textbf{d} & \textbf{e} & \textbf{f} & \textbf{g} & \textbf{h} \\
\cmidrule{2-9}
\vspace{3pt}\textbf{Grid Spacing ($\delta$)} & 27.8 m & 23.1 m & 24.8 m & 21.0 m & 24.1 m & 21.7 m & 21.7 m & 31.6 m \\
\vspace{3pt}\makecell[l]{\textbf{Number of} \\ \textbf{Candidate Waypoints ($\boldsymbol{|W|}$)}} & 4 & 5 & 7 & 10 & 5 & 7 & 7 & 10 \\
\vspace{3pt}\textbf{Selected Waypoints} & \{1,2,3,4\} & \{1,2,3,4,5\} & \{1,2,4,6,7\} & \{1,2,5,6,7,8,10\} & \{1,3,4\} & \{1,2,4,6,7\} & \{2,3,4,6,7\} & \{1,2,3,5,6,8,9\} \\
\vspace{3pt}\makecell[l]{\textbf{Number}\\ \textbf{of Selected Waypoints}} & 4 & 5 & 5 & 7 & 3 & 5 & 5 & 7 \\
\vspace{1pt}\makecell[l]{\textbf{Minimal Number}\\ \textbf{of Selected Waypoints}} & 4 & 5 & 5 & 7 & 3 & 5 & 5 & 7 \\
\bottomrule
\end{tabular}
\label{tab:EmulatorResult}
\end{table*}

The results obtained from the noise-free quantum simulator are summarized in Table~\ref{tab:EmulatorResult}, which reports the grid spacing, the number of candidate waypoints, the minimum number of waypoints required for 3-coverage, and the waypoints selected by the QAOA-based solution.
To remain within the qubit budget while preserving feasibility, the grid spacing was adjusted on a per-scenario basis.
Importantly, the proposed formulation depends only on the relative scaling between the communication radius $r$ and the grid spacing $\delta$, and thus can be uniformly scaled without loss of generality.

In every scenario, the noise-free quantum simulator consistently returned solutions that were both feasible and optimal---that is, each target grid point was covered by at least three selected waypoints, and the total number of selected waypoints matched the known minimum for 3-coverage.
These results confirm the feasibility and optimality of the proposed quantum-based UAV waypoint optimization framework under ideal (noise-free) conditions, establishing a strong baseline for subsequent evaluation on real quantum hardware.

Appendix~C of the supplementary material further examines QAOA convergence, run-to-run stability, calibration-derived quantum noise, and sensitivity to the principal QUBO and QAOA settings using scenario~(g). Because BoS solutions may remain unchanged despite variations in the sampled distribution, we additionally evaluate $P_{\mathrm{feas}}$ and $P_{\mathrm{opt}}$, denoting the fractions of feasible and globally minimum-cardinality samples, respectively. COBYLA satisfied its stopping criterion after 48 objective-function evaluations, while repeated executions showed stable BoS solution quality. Increasing the shot count mainly reduced run-to-run dispersion, whereas stronger noise reduced $P_{\mathrm{feas}}$ and $P_{\mathrm{opt}}$ without changing the retained BoS solution. The penalty analysis identified an instance-specific boundary of $\lambda_{\mathrm{crit}}=0.5$: values below it can favor infeasible assignments, whereas all tested values above it yielded feasible minimum-cardinality BoS solutions. Sensitivity to QAOA depth, optimizer, and shot count was observed primarily in the sampled distribution rather than in retained solution quality.

\section{Evaluation on Real Quantum Hardware} \label{subsec:EvaluationRealQuantum}
We evaluated the proposed QAOA-based UAV waypoint optimization framework on IBM’s 127-qubit Eagle quantum processor (\texttt{ibm\_yonsei}) using the Qiskit Runtime environment.
In contrast to the noise-free simulator, real quantum hardware introduces imperfections such as gate errors, decoherence, and crosstalk, all of which can affect solution quality.
The quantum circuit was executed using the Qiskit Runtime
\texttt{SamplerV2} with a QAOA depth of $p=2$, 1024 shots per objective-function evaluation, and COBYLA employed under the same one-evaluation budget.
For each scenario, we performed ten independent runs and averaged the results.
We assess the proposed method along three dimensions: coverage satisfaction, mission efficiency relative to existing SAR waypoint-deployment methods, and waypoint-selection quality relative to standard classical solvers.

\subsection{Coverage satisfaction}
\label{subsec:Coverage_satisfaction}

\begin{figure}[tbp]
    \centering
    \includegraphics[width=0.8\linewidth]{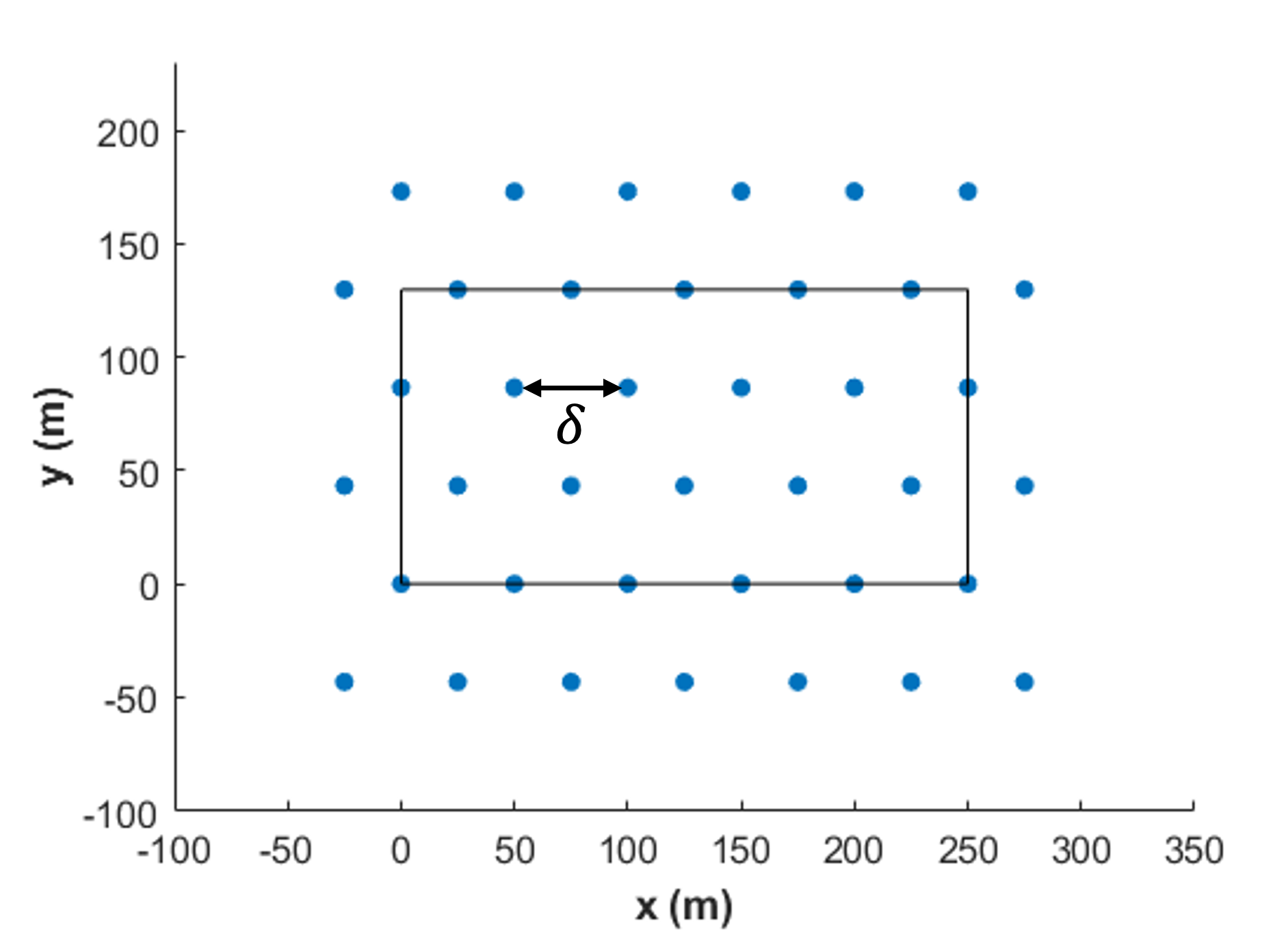}
    \caption{Example rectangular target area used in the coverage-satisfaction experiment. The target region has a physical size of 250~m~$\times$~130~m and is discretized using a triangular lattice with a grid spacing of 50~m. Lattice points are generated to cover the rectangular region, with one additional layer of points surrounding its boundary. For this example, the resulting numbers of candidate UAV waypoints and target grid points are $|\boldsymbol{W}| = 39$ and $|\boldsymbol{G}| = 22$, respectively.}
    \label{fig:SquareScenario}
\end{figure}

We first investigated whether the proposed method could consistently achieve 3-coverage as the size of a target area increases. For this purpose, we constructed a series of rectangular test scenarios with increasing area sizes. 
An example scenario, corresponding to a 250 m $\times$ 130 m target region, is shown in Fig.~\ref{fig:SquareScenario}.
Unlike in the simulator-based experiments, the 127-qubit quantum computer enables the evaluation of larger problem instances, allowing us to adopt more realistic UAV communication and flight parameters.
In these experiments, the UAV’s communication range was set to 100~m and the flight altitude was assumed to be 80~m, resulting in a 2D ground-projected coverage radius $r$ of 60~m.
The assumed 100~m communication range is consistent with prior UAV-based Wi-Fi detection results reporting ranges up to approximately 200~m \cite{Wang13:Feasibility}, while the 80~m flight altitude remains below the 400~ft (approximately 120~m) ceiling under FAA Part~107 \cite{FAA_Part107}.

Each target region was discretized using a triangular lattice with a fixed grid spacing of $\delta = 50$~m. Lattice points were generated only in the vicinity of the rectangular target region, as shown in Fig.~\ref{fig:SquareScenario}. Specifically, all lattice points lying inside the rectangle, together with one additional layer of points surrounding its boundary, were retained. This boundary margin prevents coverage degradation near the edges of the target region due to lattice truncation. All retained lattice points were considered as candidate UAV waypoints and included in the set $\boldsymbol{W}$, whereas only those lattice points lying inside the rectangular target region were included in the set of potential target locations $\boldsymbol{G}$.

\begin{figure}[htbp]
    \centering
    \includegraphics[width=0.95\linewidth]{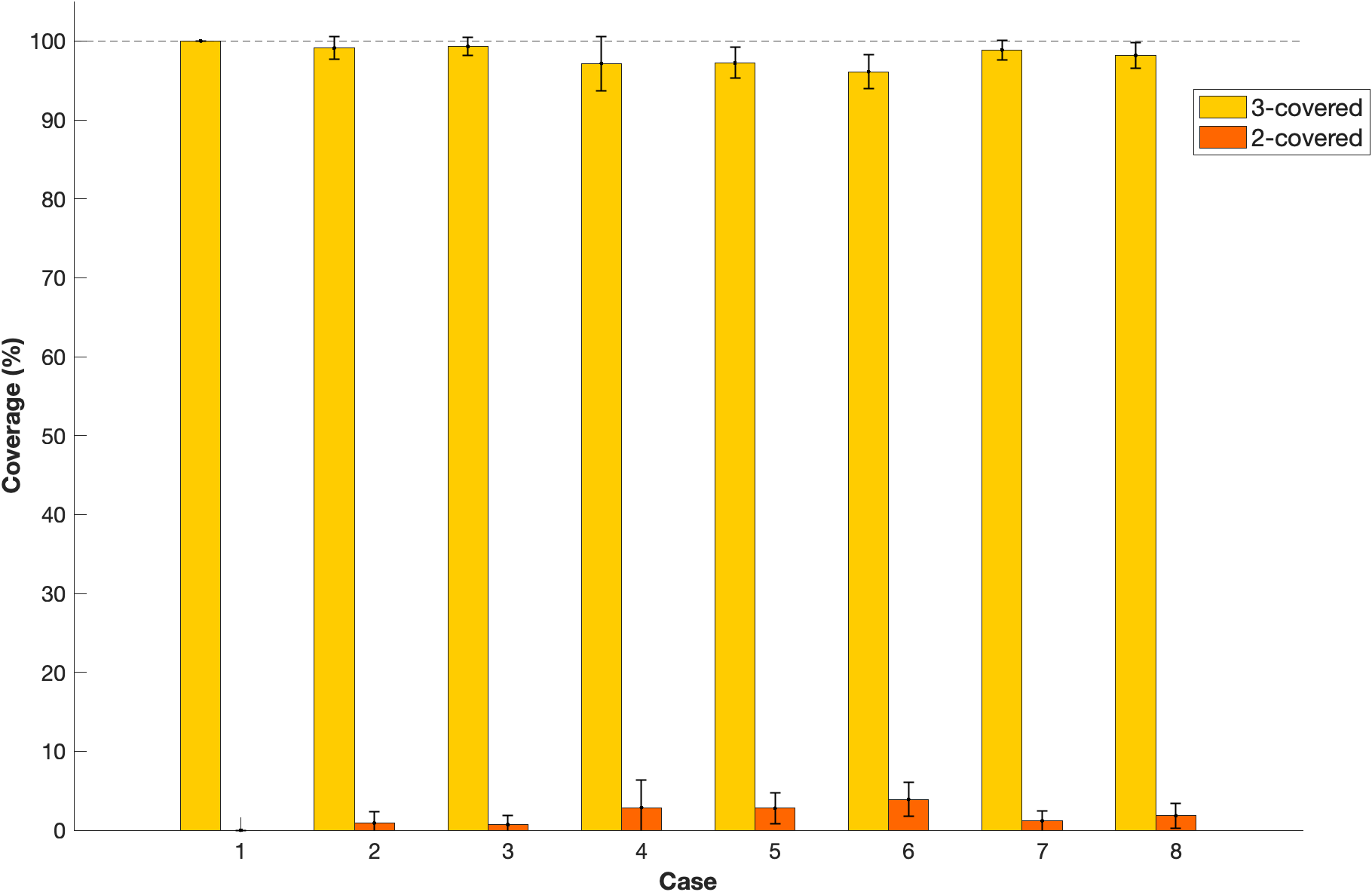}
    \caption{Coverage satisfaction (3-coverage ratio) across different test cases on real quantum hardware. Each case corresponds to a rectangular target region defined in Table~\ref{tab:ScenarioSummary}.  The QAOA-based method is executed ten times on IBM’s Eagle processor, and error bars denote one standard deviation of the measured coverage ratio. All scenarios achieve over 95\% 3-coverage despite hardware noise.}
    \label{fig:CoverageSatisfaction}
\end{figure}

Fig.~\ref{fig:CoverageSatisfaction} summarizes the coverage satisfaction achieved by the proposed method on real quantum hardware across different test cases. Each case corresponds to a rectangular target region with a distinct physical size. The considered cases span target area sizes ranging from 100~m~$\times$~87~m to 250~m~$\times$~130~m, resulting in 8 to 23 target grid points ($|\boldsymbol{G}|$) and 18 to 39 candidate UAV waypoints ($|\boldsymbol{W}|$), depending on the scenario. The detailed correspondence between the physical target region sizes and the resulting numbers of target grid points and candidate UAV waypoints is summarized in Table~\ref{tab:ScenarioSummary}.
For each target size, we executed the algorithm ten times on IBM’s Eagle processor and report the mean coverage ratio; error bars represent one standard deviation around the mean.
Across all scenarios, the proposed method achieved over 95\% 3-coverage. These results indicate that the method consistently achieves high 3-coverage ratios, even under hardware-induced noise.
The remaining target points were predominantly 2-covered, and notably, no target point was 1-covered or uncovered.

The computational resources required by the QAOA implementation depend on the number of encoded binary variables and nonzero QUBO terms. Under the fixed-density conditions considered in this work, these quantities exhibit approximately linear growth with the number of candidate waypoints. The corresponding execution time and quantum-resource requirements are analyzed in Sections~\ref{sec:RuntimeAnalysis} and~\ref{sec:QuantumResourceRequirements}.

\begin{table}[tbp]
\centering
\renewcommand{\arraystretch}{1.15}
\caption{
Summary of rectangular target areas used in the coverage-satisfaction
experiment and the corresponding numbers of grid points.
}
\label{tab:ScenarioSummary}
\begin{tabular}{
>{\centering\arraybackslash}p{0.6cm}
>{\centering\arraybackslash}p{2.4cm}
>{\centering\arraybackslash}p{1.7cm}
>{\centering\arraybackslash}p{2.2cm}
}
\toprule
\textbf{Case} &
\textbf{Target Area Size} &
\makecell{\textbf{Number of} \\ \textbf{Target} \\ \textbf{Grids} ($|\boldsymbol{G}|$)} &
\makecell{\textbf{Number of} \\ \textbf{Candidate}\\ \textbf{Waypoints} ($|\boldsymbol{W}|$)} \\
\midrule
1 & 100 m $\times$ 87 m  & 8  & 18 \\
2 & 150 m $\times$ 87 m  & 11 & 23 \\
3 & 150 m $\times$ 130 m & 14 & 27 \\
4 & 200 m $\times$ 87 m  & 14 & 28 \\
5 & 200 m $\times$ 130 m & 18 & 33 \\
6 & 200 m $\times$ 173 m & 23 & 39 \\
7 & 250 m $\times$ 87 m  & 17 & 33 \\
8 & 250 m $\times$ 130 m & 22 & 39 \\
\bottomrule
\end{tabular}
\end{table}

\subsection{Mission efficiency versus SAR deployment methods}
\label{subsec:SARDeployment}

We evaluate the proposed framework against representative SAR waypoint-deployment methods---LocalizerBee by Perazzo et al.~\cite{Perazzo16:Drone}, the method of Yuan et al.~\cite{Yuan22:A}, and a learning-based PPO-RL baseline. Each method produces or selects a waypoint set. For a fair application-level comparison, each waypoint set is converted into a closed mission route using the same TSP post-processing procedure. We then compare the number of selected waypoints and the resulting flight-path length. 
A test scenario was constructed based on the building layout of Yonsei University's International Campus.
For each method, the selected waypoints were connected using the Chained Lin-Kernighan heuristic~\cite{Applegate03:Chained} to compute a short
visiting tour.
In this evaluation, the communication radius was set to $l=150$~m and the UAV flight altitude was set to $h=80$~m, both of which remain within the realistic operational ranges discussed earlier.

\noindent\textbf{Classical Deterministic Grid-Based Baselines.}
We compared our approach against two baseline methods: LocalizerBee proposed by Perazzo et al.~\cite{Perazzo16:Drone} and the method proposed by Yuan et al.~\cite{Yuan22:A}.
LocalizerBee \cite{Perazzo16:Drone} constructs a triangular waypoint grid by tessellating the target area with isosceles triangles whose base and height are calculated such that all triangle sides remain below the maximal admissible length $L_{\max}$, which is derived from the communication range and a $\gamma_\mathrm{W}$-wide control-error margin ($\gamma_\mathrm{W}=10$~m in this paper, following prior work).
To avoid boundary coverage gaps, LocalizerBee additionally places waypoints corresponding to halved triangles along the target-region boundaries, resulting in waypoint locations at the midpoints of some triangle bases.
Under our experimental settings, the resulting triangular grid has a base length of 75.46~m and a height of 74.90~m.
Yuan et al.~\cite{Yuan22:A} generate waypoints by tiling the deployment area with squares whose side length $\rho = (\sqrt{2}/2)\, r$ is determined by the UAV’s ground-projected coverage radius $r$, ensuring that the union of all squares fully covers the deployment area.
Under the same settings, the resulting square grid has a spacing of 89.72~m between adjacent waypoints.

A fundamental difference between these prior approaches and the proposed framework lies in how waypoints are generated and utilized.
Existing methods directly use the constructed geometric grids as the final set of UAV waypoints to guarantee 3-coverage of the target area, which does not explicitly optimize the selection of a minimal subset of waypoints.
In contrast, our approach first constructs a finer-grained triangular lattice and then explicitly formulates waypoint selection as an optimization problem, selecting only a subset of lattice points that collectively achieves 3-coverage.
In this study, the lattice spacing was set to $\delta=55$~m, which is finer than the effective grid spacings used in the baseline methods.
This value represents the finest spatial resolution achievable under the available qubit constraints.

\noindent\textbf{Learning-Based Baseline.}
In addition to the classical deterministic grid-based baselines, we evaluate the proposed method against a learning-based baseline implemented using PPO \cite{Schulman17:Proximal} with action masking, hereafter referred to as PPO-RL.
Both the proposed QAOA framework and PPO-RL operate on the same set of candidate waypoints, generated by discretizing the UAV-navigable area $\mathcal{U}$ into a triangular lattice with grid spacing $\delta = 55$~m, as described in Section~\ref{Sec:Method}.
For this mission-level comparison, the RL agent is trained in a \emph{cross-domain} setting: it is trained on building footprints from two geographically distinct campuses---Yonsei University Seoul Campus (South Korea) and the University of Waterloo (Canada)---and evaluated on Yonsei University International Campus, which was entirely unseen during training. This cross-campus generalization setting reflects realistic deployment conditions in which the target environment is unknown a priori. Details of the Markov decision process (MDP) formulation, network architecture, hyperparameters, and training convergence are provided in Appendix~D of the supplementary material.

\noindent\textbf{Evaluation Protocol.}
Since both QAOA and PPO-RL involve stochastic elements---quantum measurement noise in the former and stochastic action sampling in the latter---we adopt a \emph{best-of-10} evaluation protocol for both methods: each method is executed ten independent times, and the feasible run (i.e., achieving 100\% 3-coverage) with the shortest path is selected for path comparison.
The classical deterministic grid-based baselines are deterministic and always produce feasible solutions; they are therefore evaluated with a single run.
Coverage and feasibility statistics are additionally reported across all ten runs to characterize solution reliability.

\begin{figure}[tbp]
    \centering
    \includegraphics[width=0.85\linewidth]{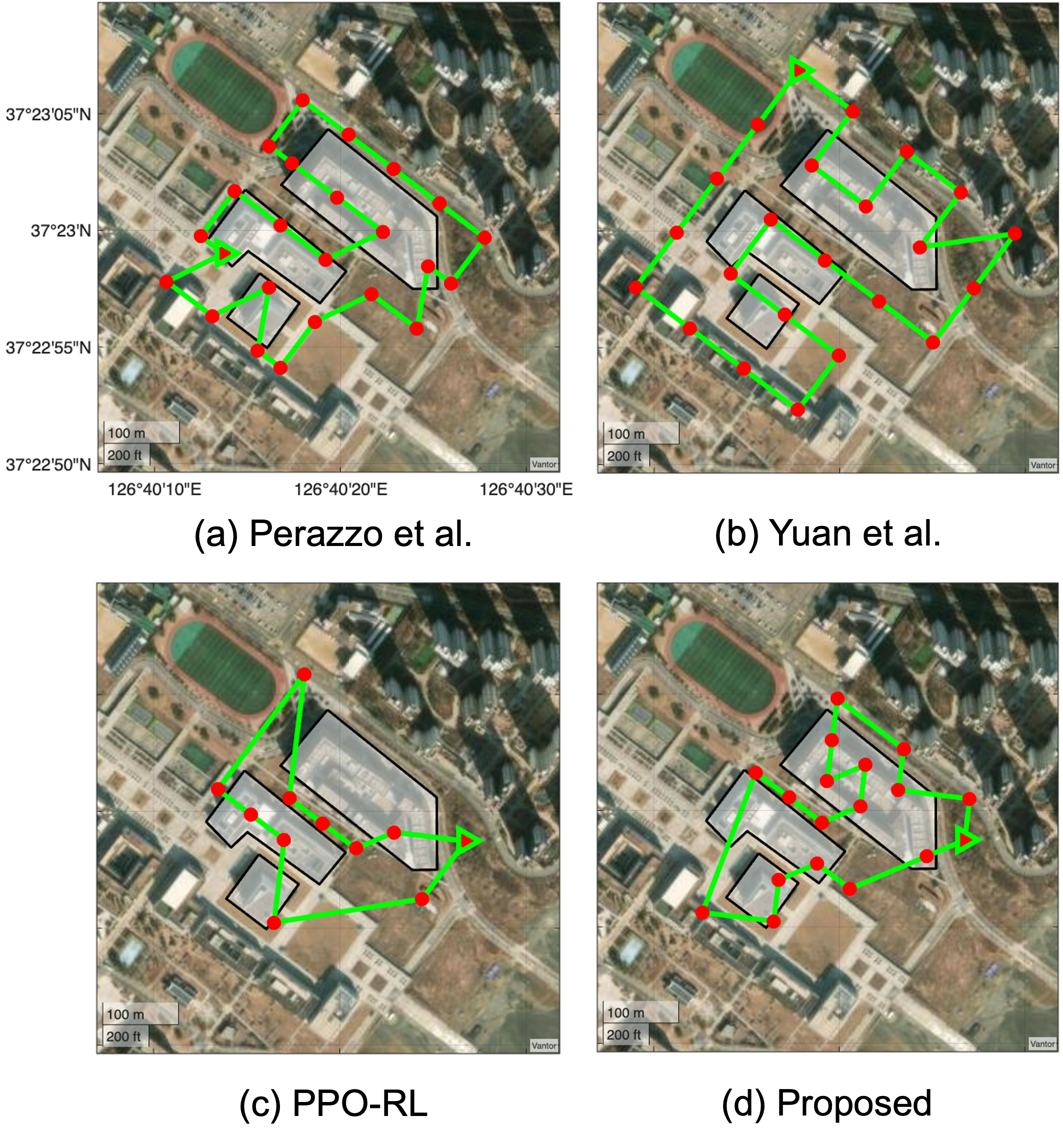}
    \caption{
Comparison of UAV waypoints and TSP-based flight paths in a realistic scenario
based on Yonsei University's International Campus:
(a)~Perazzo et al.~(LocalizerBee)~\cite{Perazzo16:Drone},
(b)~Yuan et al.~\cite{Yuan22:A},
(c)~PPO-RL baseline, and
(d)~the proposed method.
Target areas correspond to three irregular building footprints, shown as
semi-transparent white regions with black boundaries.
Red circles denote selected UAV waypoints, green lines represent the computed
flight path, and green triangles indicate the start/end location.
}\label{fig:YonseiResult}
\end{figure}

\noindent\textbf{Results.}
Fig.~\ref{fig:YonseiResult} shows the waypoints and the corresponding UAV paths for all four methods.
The target areas are defined as three irregular regions corresponding to building footprints, highlighted using semi-transparent white overlays with black boundaries.
Red circles indicate the selected UAV waypoints, green lines represent the UAV flight paths, and green triangles mark the common start/end location.
Table~\ref{tab:YonseiResultTable} summarizes the results across all methods.

The proposed QAOA framework reduces the total path length by 22.5\% compared to Perazzo et al.~(LocalizerBee) and by 37.0\% compared to Yuan et al., while also selecting fewer waypoints.
Compared to the cross-domain PPO-RL baseline, QAOA selects more waypoints (18 vs.\ 11) and produces a longer path (1323.83~m vs.\ 1129.36~m). PPO-RL and QAOA, however, represent different solution paradigms: PPO-RL applies a policy trained on other environments, whereas QAOA performs per-instance optimization using the environment model available at evaluation time. Their coverage reliability, training requirements, and evaluation assumptions differ substantially and are examined at the solver level in Section~\ref{subsec:OptLearnComparison}.

\begin{table*}[tbp]
\centering
\caption{Comparison of all methods in the Yonsei University International Campus scenario. Waypoint count and path length are reported for the best feasible run out of 10 independent runs for QAOA and PPO-RL.
For these stochastic methods, the 3-coverage ratio is reported as the mean $\pm$ standard deviation over 10 runs, and the feasible rate is also computed over the same runs. The classical grid-based methods are deterministic and were executed once; therefore, no run-to-run standard deviation is reported for these methods$^\dagger$.}
\label{tab:YonseiResultTable}
\begin{tabular}{
    p{2.4cm}
    >{\centering\arraybackslash}p{2.1cm}
    >{\centering\arraybackslash}p{2.4cm}
    >{\centering\arraybackslash}p{2.7cm}
    >{\centering\arraybackslash}p{2.4cm}
    >{\centering\arraybackslash}p{2.4cm}}
\toprule
\textbf{Method}
& \makecell{\textbf{Number of} \\ \textbf{Waypoints}}
& \makecell{\textbf{Path} \\ \textbf{Length (m)}}
& \makecell{\textbf{3-Cov. Ratio} \\ \textbf{Mean $\pm$ Std.}}
& \makecell{\textbf{Feasible} \\ \textbf{Rate}}
& \makecell{\textbf{Training} \\ \textbf{Required}} \\
\midrule
Perazzo et al.~\cite{Perazzo16:Drone}
& 24
& 1708.29
& 100\%$^\dagger$
& 100\%$^\dagger$
& No \\

Yuan et al.~\cite{Yuan22:A}
& 23
& 2100.77
& 100\%$^\dagger$
& 100\%$^\dagger$
& No \\

PPO-RL
& 11
& 1129.36
& 97.1\%$\pm$3.5\%
& 60\%
& Yes \\

\textbf{Proposed}
& \textbf{18}
& \textbf{1323.83}
& \textbf{99.3\%$\pm$2.1\%}
& \textbf{90\%}
& \textbf{No} \\
\bottomrule
\multicolumn{6}{l}{\footnotesize
$^\dagger$Classical grid-based methods guarantee full 3-coverage by
construction and are therefore always feasible.}
\end{tabular}
\end{table*}

\subsection{Waypoint-selection quality against classical solvers}
\label{subsec:OptLearnComparison}

Having compared mission efficiency against methods that produce a full flight route, we now turn to solvers that address only the waypoint-selection subproblem, in order to position the solution quality and runtime of QAOA relative to exact, heuristic, metaheuristic, and learning-based optimizers. All methods are evaluated on the same Yonsei University International Campus scenario, with $|\boldsymbol{W}| = 39$ candidate waypoints and $|\boldsymbol{G}| = 14$ target grid points. Unlike the deployment methods above, these solvers optimize only the selected waypoint subset and do not prescribe a route; their flight-path length is therefore a byproduct of the subsequent TSP stage rather than a property of the solver, and we accordingly compare them on selection quality alone. We report the number of selected waypoints, the mean 3-coverage ratio, the feasible rate, the runtime, and the approximation ratio, defined as the number of selected waypoints divided by the ILP optimum (equal to 1 for an optimal solution).

\noindent\textbf{Baselines.} 
The considered baselines comprise three families. First, the exact baseline solves the constrained problem in Eq.~\eqref{eq:ConstrainedOptimization} as an ILP, directly minimizing the number of selected waypoints subject to the $k$-coverage constraints; its optimal waypoint count serves as the denominator of the approximation ratio for all other methods. Second, the greedy, simulated annealing (SA)~\cite{Kirkpatrick83:Optimization}, and genetic algorithm (GA)~\cite{Goldberg89:Genetic} baselines represent heuristic and metaheuristic solvers. The greedy algorithm operates directly on the constrained formulation as a fast, deterministic approximation reference, whereas SA and GA optimize the same QUBO formulation in Eq.~\eqref{eqn:QUBO_final} used to construct the proposed QAOA framework,
providing classical QUBO-based references for comparison with the quantum solver. Third, PPO-RL represents a learning-based approach, evaluated under both the cross-domain regime introduced in Section~\ref{subsec:SARDeployment} and an \emph{in-domain} regime, in which the agent is additionally trained on the test campus itself, so that its training distribution includes the evaluation environment. Full implementation details of the ILP, greedy, SA, and GA baselines are provided in Appendix~E of the supplementary material, and those of PPO-RL in Appendix~D. For each stochastic method---QAOA, PPO-RL, SA, and GA---the reported waypoint count is the minimum among the feasible solutions obtained over ten independent runs, while the coverage and feasible-rate statistics are computed over all ten runs.

\noindent\textbf{Results.} Table~\ref{tab:OptLearnComparison} summarizes the results. At the problem scale considered here, the ILP baseline returns the optimal seven-waypoint solution in $0.04$~s, and the greedy, QUBO-SA, and QUBO-GA baselines each obtain an eight-waypoint feasible solution (approximation ratio $1.143$), with runtimes ranging from under $0.01$~s for the greedy method to $30.34$~s for QUBO-SA. The proposed QAOA framework returns an 18-waypoint feasible solution in approximately $30.5$~s. On instances of this size, then, mature classical solvers are both faster and more accurate, as expected on present-day NISQ hardware. What the proposed framework contributes is orthogonal to this comparison: it is the first formulation of the SAR waypoint-selection problem as an exact-penalty QUBO and its solution on real gate-based quantum hardware, in a solver-independent form that the QUBO-SA and QUBO-GA baselines share directly. The measured runtime trend over the tested range is analyzed in Section~\ref{sec:RuntimeAnalysis} as an executability indicator rather than as an asymptotic advantage over classical optimization.

The learning-based comparison is likewise informative. Under the cross-domain setting, QAOA achieves a feasible rate of $90\%$ and a mean 3-coverage ratio of $99.3\%$, whereas PPO-RL achieves $60\%$ and $97.1\%$, respectively, indicating greater run-to-run coverage reliability for the training-free QAOA in this setting. In life-critical SAR missions, incomplete coverage may leave target locations under-covered and reduce the number of usable measurements for localization, making robust coverage planning operationally relevant. The two methods, however, operate under different paradigms---PPO-RL applies a policy learned from other environments, whereas QAOA performs per-instance optimization---so these figures reflect a trade-off between cross-environment generalization and training-free per-instance optimization rather than a general superiority of either method.

The in-domain PPO regime further clarifies this trade-off. Because the exact evaluation instance is included during training, its result should be interpreted as a favorable in-domain sensitivity reference rather than as a held-out generalization result. Under this setting, PPO-RL reaches the ILP-optimal count of seven waypoints and a feasible rate of 90\%. Taken together, the cross-domain and in-domain results indicate that PPO-RL benefits substantially from environment-specific training, whereas QAOA performs per-instance optimization without prior policy training. QAOA nevertheless requires the environment and coverage model to be available at optimization time; therefore, these results characterize different operating paradigms rather than establishing a general superiority of either method.

\begin{table*}[tbp]
\centering
\caption{Solver-level comparison of waypoint-selection quality on the Yonsei University International Campus scenario ($|\boldsymbol{W}|=39$). For stochastic methods, the waypoint count is the best feasible value over 10 runs, while the 3-coverage ratio is reported as mean $\pm$ standard deviation and the feasible rate is computed over all runs. ILP and greedy are deterministic and are reported from a single run. The approximation ratio is relative to the ILP optimum.}

\label{tab:OptLearnComparison}
\begin{tabular}{
    l
    >{\centering\arraybackslash}p{2.0cm}
    >{\centering\arraybackslash}p{2.5cm}
    >{\centering\arraybackslash}p{1.6cm}
    >{\centering\arraybackslash}p{2.0cm}
    >{\centering\arraybackslash}p{1.6cm}
    >{\centering\arraybackslash}p{1.8cm}}
\toprule
\textbf{Method}
& \makecell{\textbf{Number of}\\\textbf{Waypoints}}
& \makecell{\textbf{3-Cov. Ratio}\\\textbf{Mean $\pm$ Std.}}
& \makecell{\textbf{Feasible}\\\textbf{Rate}}
& \textbf{Runtime}
& \makecell{\textbf{Approx.}\\\textbf{Ratio}}
& \makecell{\textbf{Training}\\\textbf{Required}} \\
\midrule

\multicolumn{7}{l}{\emph{Exact}}\\
ILP
& 7
& 100\%
& 100\%
& 0.04~s
& 1.000
& No \\

\midrule
\multicolumn{7}{l}{\emph{Approximation / metaheuristic}}\\
Greedy
& 8
& 100\%
& 100\%
& $<0.01$~s
& 1.143
& No \\

QUBO-SA
& 8
& 100.0\%$\pm$0.0\%
& 100\%
& 30.34~s
& 1.143
& No \\

QUBO-GA
& 8
& 100.0\%$\pm$0.0\%
& 100\%
& 3.41~s
& 1.143
& No \\

\midrule
\multicolumn{7}{l}{\emph{Learning-based}}\\
PPO-RL (cross-domain)
& 11
& 97.1\%$\pm$3.5\%
& 60\%
& $\approx$12~ms$^\dagger$
& 1.571
& Yes \\

PPO-RL (in-domain)
& 7
& 99.3\%$\pm$2.1\%
& 90\%
& $\approx$12~ms$^\dagger$
& 1.000
& Yes \\

\midrule
\multicolumn{7}{l}{\emph{Proposed}}\\
\textbf{QAOA}
& \textbf{18}
& \textbf{99.3\%$\pm$2.1\%}
& \textbf{90\%}
& \textbf{$\approx$30.5~s}
& \textbf{2.571}
& \textbf{No} \\

\bottomrule
\multicolumn{7}{l}{\footnotesize
$^\dagger$Inference time only; PPO-RL additionally requires
${\approx}312.9$~s (cross-domain) or ${\approx}296.7$~s (in-domain)
of offline training.}
\end{tabular}
\end{table*}

\section{Computational Runtime and Complexity}
\label{sec:RuntimeAnalysis}

\subsection{Measured runtime}

Table~\ref{tab:runtime} summarizes the measured runtimes of all methods for the Yonsei University International Campus scenario ($|\boldsymbol{W}|=39$). The classical deterministic grid-based baselines of Perazzo et al.~\cite{Perazzo16:Drone} and Yuan et al.~\cite{Yuan22:A} complete within milliseconds, as they directly construct waypoint grids without any optimization step. Among the optimization-based classical methods, the ILP baseline completes in approximately 0.04~s, while the runtimes of the greedy, QUBO-SA, and QUBO-GA baselines range from less than 10~ms to approximately 30.34~s on the classical workstation. The PPO-RL baseline requires approximately 312.9~s (cross-domain) or 296.7~s (in-domain) of offline training, after which inference at test time requires only approximately 12~ms per instance. For the proposed framework, the end-to-end runtime for this scenario consists of triangular-lattice generation ($\approx$ 34~ms), coverage set construction (computing $\boldsymbol{N}_j$ for all $j \in \boldsymbol{G}$, $\approx$ 96~ms), and QAOA-based optimization. The two classical preprocessing stages together require approximately 130~ms and are negligible compared to the QAOA execution time. The QAOA execution time is reported as approximately 30.5~s, based on the measured value for the rectangular test instance of the same size ($|\boldsymbol{W}|=39$); the reported value corresponds to measured circuit execution time on IBM's 127-qubit Eagle processor, excluding cloud queue latency. All classical method runtimes were measured on an Intel Core i7-10700K processor with 64~GB RAM.

\begin{table}[tbp]
\centering
\caption{Measured runtimes for the Yonsei University International Campus scenario ($|\boldsymbol{W}|=39$). The QAOA execution time is based on the rectangular test instance of the same size ($|\boldsymbol{W}|=39$); see Fig.~\ref{fig:runtime}.}
\label{tab:runtime}
\begin{tabular}{p{2.6cm} >{\raggedright\arraybackslash}p{4.8cm}}
\toprule
\textbf{Method} & \textbf{Measured Runtime} \\
\midrule
Perazzo et al.~\cite{Perazzo16:Drone} & $<10$~ms \\
Yuan et al.~\cite{Yuan22:A}           & $<10$~ms \\
ILP                               & $\approx$ 0.04~s \\
Greedy                            & $<10$~ms \\
QUBO-SA                           & $\approx$ 30.34~s \\
QUBO-GA                           & $\approx$ 3.41~s \\
PPO-RL                                 & $\approx$ 312.9~/~296.7~s (training) / $\approx$ 12~ms (inference) \\
Proposed                               & $\approx$ 30.5~s \\
\bottomrule
\end{tabular}
\end{table}

\begin{figure}[tbp]
    \centering
    \includegraphics[width=0.8\linewidth]{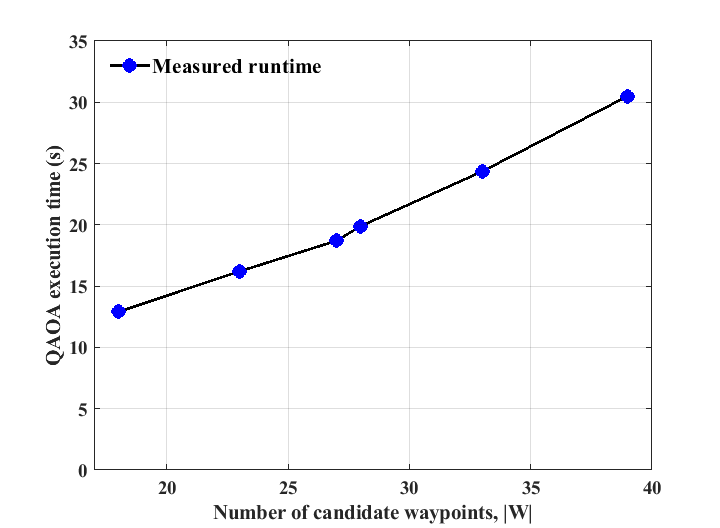}
    \caption{
Measured QAOA execution time as a function of the number of candidate waypoints $|\boldsymbol{W}|$,
evaluated on IBM's 127-qubit Eagle processor over the rectangular test instances in Table~\ref{tab:ScenarioSummary}.
For $|\boldsymbol{W}| = 33$ and $39$, the plotted value is the mean over the two corresponding cases. The runtime exhibits an approximately linear trend as $|\boldsymbol{W}|$ increases from 18 to 39.
}
    \label{fig:runtime}
\end{figure}

Fig.~\ref{fig:runtime} plots the measured QAOA execution time against the number of candidate waypoints $|\boldsymbol{W}|$ for the rectangular test instances summarized in Table~\ref{tab:ScenarioSummary}.
As $|\boldsymbol{W}|$ increases from 18 to 39, the QAOA execution time grows from 12.9~s to 30.5~s, exhibiting an approximately linear trend.
Note that for $|\boldsymbol{W}| = 33$ and $|\boldsymbol{W}| = 39$, two distinct target area sizes exist in Table~\ref{tab:ScenarioSummary}; the plotted values represent the mean execution time averaged over the two corresponding cases.
As described in Section \ref{Sec:QuantumOptimization}, the classical optimizer iteratively updates the QAOA parameters by repeatedly estimating the expected energy in Eq. \eqref{eq:qaoa-energy}. Let $N_{\mathrm{eval}}$ denote the number of such objective-function evaluations, and let $S$ denote the shot count per evaluation. Since each evaluation requires repeated executions of the depth-$p$ QAOA circuit, the dominant runtime of the QAOA stage can be approximated as
\begin{equation}
T_{\mathrm{QAOA}} \approx N_{\mathrm{eval}}\, S\, \tau_{\mathrm{circ}}(n,D_{\mathrm{total}}),
\label{eq:tqaoa}
\end{equation}
where $\tau_{\mathrm{circ}}(n,D_{\mathrm{total}})$ denotes the execution cost of a single $n$-qubit circuit of depth $D_{\mathrm{total}}$, $n = |\boldsymbol{W}| + \sum_{j \in G} B_j$ is the total qubit count, and $D_{\mathrm{total}} = \Theta(p\Delta^{\star}) + O(p)$ is the circuit depth. A detailed discussion of the circuit depth $D_{\mathrm{total}}$ and the associated quantum resource requirements is provided in Section~\ref{sec:QuantumResourceRequirements}. Under the fixed-density conditions discussed in the complexity analysis, $n$ and $M_Q$ can scale as $O(|\boldsymbol{W}|)$, which characterizes the growth of the QUBO representation and logical gate count under those conditions. This does not imply linear end-to-end QAOA runtime, because $N_{\mathrm{eval}}$, the sampling effort, hardware-routing overhead, and the compiled circuit depth may also vary with problem size. Accordingly, the approximately linear increase observed in Fig.~\ref{fig:runtime} should be interpreted as an empirical trend over the tested range of $|\boldsymbol{W}|=18$--$39$, rather than as a general asymptotic scaling result.
 
Note that these runtimes were obtained on heterogeneous computational platforms and correspond to different computational stages; they therefore serve as practical reference values rather than a controlled runtime comparison. In particular, the deterministic grid-based methods perform direct waypoint construction, PPO-RL separates offline training from low-latency inference, the classical optimization baselines execute on a conventional workstation, and the QAOA value reflects gate-based quantum circuit execution.

\subsection{Complexity analysis}
\label{subsec:ComplexityAnalysis}

We next analyze the computational characteristics of the classical preprocessing and quantum optimization stages. The preprocessing phase constructs the spatial discretization, waypoint--target coverage relation, and QUBO coefficients; the same QUBO construction is also used by the QUBO-SA and QUBO-GA baselines.

\noindent\textbf{Preprocessing and QUBO Construction.}
Generating a triangular lattice containing $P$ grid points requires $O(P)$ time, while a direct construction of the waypoint--target coverage relation requires $O(|\boldsymbol{W}||\boldsymbol{G}|)$ distance evaluations. Let $r_j=|\boldsymbol{N}_j|+B_j$, where $|\boldsymbol{N}_j|$ is the number of candidate waypoints covering target point $j$ and $B_j$ is the corresponding number of slack bits. The waypoint-minimization objective contributes $O(|\boldsymbol{W}|)$ linear-coefficient updates, whereas expanding the squared coverage penalty for target $j$ requires $O(r_j^2)$ coefficient updates. Hence, direct QUBO construction requires
$O\left(|\boldsymbol{W}|+\sum_{j\in\boldsymbol{G}}r_j^2\right)$ time. 
Let $M_Q$ denote the number of nonzero linear and quadratic QUBO coefficients; then
$M_Q=O\left(|\boldsymbol{W}|+\sum_{j\in\boldsymbol{G}}r_j^2\right)$, and sparse storage requires $O(n+M_Q)$ memory.

The QUBO-to-Ising transformation in Eqs.~(\ref{eq:op-mapping})--(\ref{eq:HC_final}) maps each nonzero QUBO coefficient to a constant number of identity, $Z_i$, and $Z_iZ_l$ terms. Thus, constructing the cost and mixer Hamiltonians requires $O(n+M_Q)$ time and $O(n+M_Q)$ storage when only nonzero Pauli terms are retained. Under fixed grid spacing, coverage radius, and spatial density, the local neighborhood sizes $|\boldsymbol{N}_j|$ and slack-bit counts $B_j$ remain bounded. If $|\boldsymbol{G}|=\Theta(|\boldsymbol{W}|)$, both $n$ and $M_Q$ therefore grow linearly in order with $|\boldsymbol{W}|$. This result is conditional on bounded local coverage density and does not constitute a general guarantee for arbitrary coverage graphs.

\noindent\textbf{Quantum Optimization.}
A logical QAOA layer contains one mixer operation per qubit and cost-Hamiltonian operations associated with the nonzero QUBO terms. A depth-$p$ circuit therefore contains $O(p(n+M_Q))$ logical gates. Its logical circuit depth, governed by the maximum interaction degree $\Delta^{\star}$ defined in Section~\ref{subsec:circuit_depth}, is
$D_{\mathrm{total}}=\Theta(p\Delta^{\star})+O(p)$
before hardware-specific transpilation. The end-to-end quantum optimization cost additionally includes $N_{\mathrm{eval}}S$ circuit executions, as expressed in Eq.~\eqref{eq:tqaoa}. Consequently, the conditional linear growth of $n$ and $M_Q$ does not imply linear end-to-end QAOA runtime. The approximately linear execution-time trend observed over the tested range in Fig.~\ref{fig:runtime} is therefore empirical rather than a general scaling guarantee; larger instances must additionally account for optimizer evaluations, sampling effort, hardware-routing overhead, and compiled circuit depth.

\noindent\textbf{Comparison with Classical Optimization Methods.}
Table~\ref{tab:complexity} summarizes the computational characteristics of the compared methods. The waypoint-selection problem is NP-hard, and exact ILP has exponential worst-case complexity, although the solver used in our experiments handles the tested instances efficiently. A direct greedy implementation requires $O(|\boldsymbol{W}|^2|\boldsymbol{G}|)$ time in the worst case. With direct QUBO-energy evaluation, QUBO-SA requires $O(I_{\mathrm{SA}}M_Q)$ per run and QUBO-GA requires $O(I_{\mathrm{GA}}N_{\mathrm{pop}}M_Q)$ per run. QAOA is also heuristic and provides neither a polynomial-time guarantee nor a guarantee of recovering the global optimum at finite circuit depth. Therefore, the exponential worst-case behavior of exact classical optimization should not be interpreted as evidence of improved asymptotic complexity for QAOA.

\begin{table}[tbp]
\centering
\renewcommand{\arraystretch}{1.25}
\caption{Computational characteristics of the compared optimization methods. Here, $M_Q$ is the number of nonzero QUBO coefficients, $I_{\mathrm{SA}}$ is the total number of SA moves, $I_{\mathrm{GA}}$ is the number of GA generations, and $N_{\mathrm{pop}}$ is the GA population size.}
\label{tab:complexity}
\begin{tabular}{
    >{\raggedright\arraybackslash}p{1.7cm}
    >{\raggedright\arraybackslash}p{5.6cm}}
\toprule
\textbf{Method} & \textbf{Computational Characteristic} \\
\midrule
ILP & Exponential in the worst case for exact solution \\
Greedy & $O(|\boldsymbol{W}|^2|\boldsymbol{G}|)$ for the direct implementation \\
QUBO-SA & $O(I_{\mathrm{SA}}M_Q)$ per run with direct QUBO-energy evaluation \\
QUBO-GA & $O(I_{\mathrm{GA}}N_{\mathrm{pop}}M_Q)$ per run with direct QUBO-energy evaluation \\
QAOA & $O(p(n+M_Q))$ logical gates per circuit and $N_{\mathrm{eval}}S$ circuit executions \\
\bottomrule
\end{tabular}
\end{table}

\section{Quantum Resource Requirements} \label{sec:QuantumResourceRequirements}
To assess the scalability and practical feasibility of the proposed QAOA-based optimization framework, we analyze the quantum resource requirements in terms of the number of qubits and circuit depth as a function of problem size. This analysis allows us to estimate the applicability of our approach on current and near-term quantum devices and serves as a baseline for guiding future deployment.

\subsection{Qubit estimation}
Each candidate waypoint requires one binary variable to indicate whether it is selected, and each target point requires additional slack variables to enforce the $k$-coverage constraint. Specifically, for a target point $j$, let $\boldsymbol{N}_j$ be the set of waypoints that can cover it---i.e., those within the communication radius $r$. To allow for constraint slackness, a slack variable $s_j$ is introduced with a possible range of values $0$ to $|\boldsymbol{N}_j| - k$. This requires $B_j = \lceil \log_2 (|\boldsymbol{N}_j| - k + 1) \rceil$ binary variables.

Let $|\boldsymbol{W}|$ be the number of candidate waypoints and $|\boldsymbol{G}|$ the number of target points.
The total number of qubits is therefore
\begin{equation}
\label{eq:total-qubits}
n \;=\; |\boldsymbol{W}| \;+\; \sum_{j\in \boldsymbol{G}} B_j
\;=\; |\boldsymbol{W}| \;+\; \sum_{j\in \boldsymbol{G}} 
\Big\lceil \log_2\!\big(|\boldsymbol{N}_j|-k+1\big)\Big\rceil .
\end{equation}
For reporting and sizing purposes, a practical upper bound is obtained via $N_{\max}$, where $N_{\max}=\max_{j} |\boldsymbol{N}_j|$:
\begin{equation}
\label{eq:total-qubits-ub}
n \;\le\; |\boldsymbol{W}| \;+\; |\boldsymbol{G}|\,
\Big\lceil \log_2\!\big(N_{\max}-k+1\big)\Big\rceil .
\end{equation}

\subsection{Circuit depth} \label{subsec:circuit_depth}
When the cost Hamiltonian $H_C$ is converted to the Ising form, only single-qubit $Z$ terms and two-qubit $ZZ$ interaction terms remain.
Single-qubit $Z$ terms act independently and can be applied in parallel within a single layer, contributing negligibly to circuit depth.
By contrast, $ZZ$ interactions act on two qubits simultaneously, so two distinct $ZZ$ gates cannot be applied to the same qubit at the same time.
Consequently, the depth of a single layer is proportional to the maximum number of $ZZ$ gates that must be scheduled on any qubit within the same time slot.

We refer to the qubits encoding the decision variables \(x_i\) as \emph{decision qubits} and the qubits encoding the slack bits \(c_{j,b}\) as \emph{slack qubits}.
In the cost Hamiltonian \(H_C\) (see \eqref{eq:HC_final}), the squared coverage constraint for target \(j\) involves \(Z_i\) acting on the decision qubits for \(i\in\boldsymbol{N}_j\) and
\(Z_{j,b}\) acting on the slack qubits for \(b=0,\ldots,B_j-1\).

Expanding \(\big(\sum_{i\in\boldsymbol{N}_j} Z_i\big)^2\) gives identity terms and \(ZZ\) interactions between every distinct pair \(i<i'\) in \(\boldsymbol{N}_j\):
\begin{equation}
\Bigg(\sum_{i\in \boldsymbol{N}_j} Z_i\Bigg)^{\!2}
=
\sum_{i\in \boldsymbol{N}_j} Z_i^2
\;+\;
2 \sum_{\substack{i<i'\\ i,i'\in \boldsymbol{N}_j}} Z_i Z_{i'} .
\end{equation}
Thus each decision qubit \(i\in\boldsymbol{N}_j\) is incident to \(|\boldsymbol{N}_j|-1\) distinct \(ZZ\) interactions.

Similarly, expanding \(\big(\sum_{b=0}^{B_j-1} w_b\, Z_{j,b}\big)^2\) yields identity terms and \(ZZ\) interactions between every distinct pair of slack qubits for target \(j\):
\begin{equation}
\begin{aligned}
&\Bigg(\sum_{b=0}^{B_j-1} w_b\, Z_{j,b}\Bigg)^{\!2}
= \\
&\qquad \sum_{b=0}^{B_j-1} w_b^{2}\, Z_{j,b}^{2}
\;+\;
2\sum_{\substack{b<b'\\ 0\le b,b'\le B_j-1}} w_b w_{b'}\, Z_{j,b} Z_{j,b'}.
\end{aligned}
\end{equation}
Hence each slack qubit \(Z_{j,b}\) is incident to \(B_j-1\) distinct \(ZZ\) interactions.

The cross term induces \(ZZ\) interactions between every decision qubit and every slack qubit for target \(j\):
\begin{equation}
\sum_{i\in \boldsymbol{N}_j}\sum_{b=0}^{B_j-1} w_b\, Z_i Z_{j,b}.
\end{equation}
Thus each decision qubit \(i\in\boldsymbol{N}_j\) is incident to \(B_j\) distinct \(ZZ\) interactions, and each slack qubit \(Z_{j,b}\) is incident to \(|\boldsymbol{N}_j|\).

For a fixed target \(j\), the number of distinct \(ZZ\) interactions incident on
each decision qubit \(i\in\boldsymbol{N}_j\) and on each slack qubit \(Z_{j,b}\) is
\begin{equation}
\#ZZ^{(j)}(i)=|\boldsymbol{N}_j|+B_j-1, \; \,
\#ZZ^{(j)}(j,b)=|\boldsymbol{N}_j|+B_j-1.
\end{equation}
When the same decision-qubit pair appears in multiple target constraints, the corresponding quadratic contributions are aggregated into a single $ZZ$ term. After this aggregation, let $\Delta^\star$ denote the maximum number of distinct $ZZ$ interactions incident on any logical qubit:
\begin{equation}
\label{eqn:delta}
\begin{aligned}
\Delta^\star=\max\Bigg\{&
\max_{i\in\boldsymbol{W}}
\left[
\left|
\bigcup_{j:\,i\in\boldsymbol{N}_j}
\left(\boldsymbol{N}_j\setminus\{i\}\right)
\right|
+
\sum_{j:\,i\in\boldsymbol{N}_j}B_j
\right],\\
&
\max_{j\in\boldsymbol{G}}
\left(|\boldsymbol{N}_j|+B_j-1\right)
\Bigg\}.
\end{aligned}
\end{equation}

In Eq.~(\ref{eqn:delta}), the first term counts the
distinct decision- and slack-qubit neighbors of each decision qubit, whereas the second gives the interaction degree of each slack qubit.
The aggregated terms form a simple interaction graph. 
Since interactions sharing a qubit cannot be executed simultaneously, a proper edge coloring schedules one cost layer in either $\Delta^\star$ or $\Delta^\star+1$ parallel $ZZ$ rounds.
Hence, the ideal all-to-all logical depths scale as
\begin{equation}
\label{eq:Depth}
D_{\mathrm{per\mbox{-}layer}}
=
\Theta\!\left(\Delta^\star\right),
\qquad
D_{\mathrm{total}}
=
\Theta\!\left(p\Delta^\star\right),
\end{equation}
where $D_{\mathrm{per\mbox{-}layer}}$ and
$D_{\mathrm{total}}$ denote the logical depths of one QAOA repetition and the complete depth-$p$ circuit, respectively, before hardware-specific routing and native-gate decomposition.
Constant-depth single-qubit and state-preparation layers are omitted from the asymptotic expressions.

For the Yonsei University International Campus scenario in Fig.~\ref{fig:YonseiResult}(d), the $39$ decision and $59$ slack variables yield $n=98$ logical qubits.
After aggregation, the cost Hamiltonian contains $1{,}556$ distinct $ZZ$ interactions with $\Delta^\star=85$.
An $85$-round proper edge coloring attains the lower bound imposed by $\Delta^\star$; thus, the $p=3$ circuit requires $255$ logical $ZZ$-interaction rounds.
Including one initial Hadamard layer and one single-qubit cost and mixer layer per repetition gives an ideal all-to-all logical depth of $262$, excluding measurement.
Transpilation to IBM's 127-qubit Eagle
processor (\texttt{ibm\_yonsei}) increases the compiled depth to approximately $3.1\times10^{4}$ because of restricted connectivity, routing, and native-gate decomposition.

\subsection{Large-Scale Resource Scaling}
\label{subsec:large_scale}

Large-scale formulation-level resource scaling was evaluated by expanding the triangular-lattice construction used in the coverage-satisfaction experiments of Section~\ref{subsec:Coverage_satisfaction}. The lattice spacing was fixed at $\delta=50$~m, the UAV altitude at $h=80$~m, and the three-dimensional communication range at $l=100$~m, yielding a ground-projected coverage radius of $r=60$~m. The one-layer waypoint boundary and the $k=3$ coverage requirement were retained. The target-region extent was progressively increased while preserving complete lattice rows, resulting in instances close to the nominal scales of 100, 500, 1,000, and 2,000 candidate waypoints. All retained lattice points formed the candidate set $\boldsymbol{W}$, whereas the points inside the target region formed $\boldsymbol{G}$. For each instance, the coverage constraints and slack-variable structure were generated using the same QUBO formulation as in the QAOA experiments. The resulting binary-variable count and nonzero QUBO coefficients were then evaluated without constructing or executing the corresponding large QAOA circuits.

Across the four instances, $6\leq|\boldsymbol{N}_j|\leq7$, yielding $B_j\in\{2,3\}$. The total number of binary variables is $n=|\boldsymbol{W}|+\sum_{j\in\boldsymbol{G}}B_j$, which determines the logical QAOA circuit width before hardware mapping. As defined in the preceding complexity analysis, $M_Q$ denotes the number of nonzero linear and quadratic coefficients in the resulting QUBO. Because the local coverage sizes and slack-bit counts remain bounded and $|\boldsymbol{G}|=\Theta(|\boldsymbol{W}|)$ under this fixed-density area expansion, both $n$ and $M_Q$ scale approximately linearly with $|\boldsymbol{W}|$ for these generated instances. This result characterizes the growth of the QUBO representation and circuit width under the specified geometric conditions, rather than the end-to-end runtime of QAOA.

As shown in Table~\ref{tab:large_scale_resources}, the largest generated instance contains 1,992 candidate waypoints and 1,863 target grid points, resulting in $n=7{,}535$ binary variables, including 1,992 waypoint-decision variables and 5,543 slack variables, together with 68,751 nonzero QUBO coefficients. A monolithic QAOA implementation would therefore require a circuit width of at least 7,535 qubits before any additional hardware- or fault-tolerance-related overhead. The corresponding hardware-level circuit depth and two-qubit gate count would further depend on the interaction graph, device connectivity, and transpilation. Accordingly, these values quantify formulation-level resource requirements and do not imply that such instances are executable on current gate-based quantum hardware.

\begin{table}[!t]
\caption{Large-Scale Resource Requirements Under Area Expansion}
\label{tab:large_scale_resources}
\centering
\begin{tabular}{cccc}
\hline
$|\boldsymbol{W}|$ &
$|\boldsymbol{G}|$ &
\shortstack{$n$} &
\shortstack{$M_Q$} \\
\hline
 105   & 77    & 326     & 2{,}805   \\
 492   & 429   & 1{,}757 & 15{,}789  \\
 1{,}003 & 912 & 3{,}707 & 33{,}622 \\
 1{,}992 & 1{,}863 & 7{,}535 & 68{,}751 \\
\hline
\end{tabular}
\end{table}

\section{Discussion and Future Work}

The present study considers a controlled, static, single-UAV SAR setting and demonstrates the feasibility of executing the proposed formulation on current quantum hardware. The results do not establish a quantum advantage over classical optimization: at the tested problem scale, classical exact and heuristic solvers achieve equal or better solution quality and generally shorter runtimes. The primary contribution is therefore the solver-independent QUBO formulation, its implementation on real gate-based quantum hardware, and the characterization of its current resource and solution-quality limitations.

Several practical SAR constraints can be incorporated into or coupled with the current framework. Static obstacles and no-fly zones can be handled by excluding inaccessible locations from the candidate waypoint set $\boldsymbol{W}$ through the separation of the UAV-navigable region $\mathcal{U}$ and target region $\mathcal{T}$. A waypoint-count budget, $\sum_{i\in\boldsymbol{W}} x_i \leq B_{\max},$ can provide a simple approximation of endurance limits, whereas more realistic energy and mobility models involving travel distance, hovering time, or vehicle dynamics would require joint waypoint and route optimization.

The binary coverage model also assumes sufficiently accurate knowledge of the RF coverage relation and candidate target region. In disaster environments, NLOS propagation, multipath, debris, RF interference, and communication interruptions may make coverage spatially nonuniform or stochastic, while localization or ranging uncertainty may further reduce measurement reliability. A conservative design may use a reduced effective radius, $r_{\mathrm{design}}=(1-\varepsilon)r$ ($0<\varepsilon<1$), or additional coverage redundancy, while more realistic models may replace the nominal disk model with a site-specific coverage matrix obtained from propagation simulations, radio maps, or measurements. Uncertainty in $\mathcal{T}$ may similarly be handled by enlarging the candidate target region or by updating $\mathcal{T}$ and the corresponding coverage sets as new mission information becomes available, followed by reconstruction and re-optimization of the QUBO. Robust and probabilistic formulations that explicitly model uncertain connectivity and measurement uncertainty remain important extensions.

Scalability is currently limited by both quantum hardware and the selected encoding. In the present experiments, IBM's 127-qubit Eagle processor accommodates instances with approximately 40 candidate waypoints, while the number of slack variables, QUBO interactions, and circuit depth increase with problem size and coverage connectivity. Hardware transpilation further increases the compiled depth because of restricted connectivity. More compact encodings, sparse formulations, circuit decomposition, and problem-specific mixers may reduce these requirements. Error-mitigation techniques such as measurement-error mitigation, dynamical decoupling, Pauli twirling, and zero-noise extrapolation may improve hardware performance at the cost of additional circuit executions, while hybrid quantum--classical approaches based on classical preprocessing, warm starts, or problem decomposition may extend the tractable problem scale.

Finally, multi-UAV cooperative planning could introduce UAV-specific waypoint-assignment variables together with workload, communication, collision-avoidance, and routing constraints. Dynamic targets and changing environmental information could be addressed through receding-horizon optimization, in which the environment and coverage model are repeatedly updated and re-optimized. Practical online operation, however, would depend on whether the total optimization latency remains sufficiently small relative to the rate of environmental change. These extensions therefore remain future research directions rather than capabilities demonstrated in the present work.

\section{Conclusion}

This paper presented a QAOA-based framework for UAV waypoint selection in RF-based SAR missions. We formulated the problem as an extended $k$-coverage task with independently defined UAV-navigable and target regions and derived an exact-penalty QUBO formulation with a sufficient condition guaranteeing equivalence to the original minimum-cardinality problem. The resulting formulation was evaluated using QAOA on both a noise-free simulator and IBM's 127-qubit Eagle processor. QAOA recovered the known minimum-cardinality solutions in the tested noise-free instances and achieved over 95\% 3-coverage across the real-hardware experiments. In the campus-scale scenario, the selected waypoint set also reduced the flight-path length relative to the deterministic grid-based deployment baselines. Comparisons with classical optimization and PPO-RL further characterized the solution-quality and runtime trade-offs of the proposed formulation, while the computational and quantum-resource analyses quantified its requirements from the tested hardware scale to larger generated instances. These results demonstrate the end-to-end implementation of the proposed exact QUBO formulation on current gate-based quantum hardware and provide a basis for investigating larger SAR waypoint-selection instances as quantum hardware advances.

\section*{Acknowledgment}

Generative AI was used solely to assist with grammar and language improvements during the manuscript preparation process.  
No content, ideas, data, or citations were generated by AI.  
All technical content, methodology, analysis, and conclusions were written and verified solely by the authors.

\bibliographystyle{IEEEtran}
\bibliography{mybibfile, IUS_publications}

\begin{IEEEbiography}[{\includegraphics[width=1in,height=1.25in,clip,keepaspectratio]{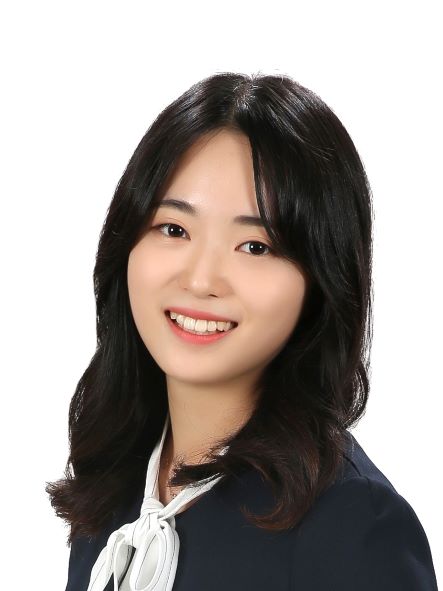}}]{Halim Lee} is a Ph.D. candidate in the School of Integrated Technology, Yonsei University, Incheon, Korea. She received the B.S. degree in Integrated Technology from Yonsei University. Her research interests include target localization and tracking, alternative PNT systems, and quantum applications for PNT.
Ms. Lee received the Undergraduate and Graduate Fellowships from the Information and Communications Technology (ICT) Consilience Creative Program supported by the Ministry of Science and ICT, Republic of Korea.
\end{IEEEbiography}

\begin{IEEEbiography}
[{\includegraphics[width=1in,height=1.25in,clip,keepaspectratio]{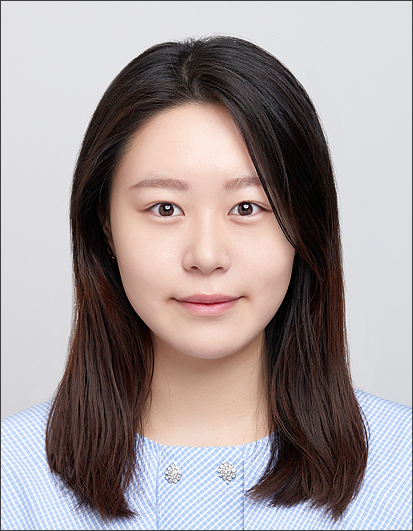}}]{Suhui Jeong}
is a Ph.D. candidate in the School of Integrated Technology, Yonsei University, Incheon, Korea. She received the B.S. degree in Integrated Technology from Yonsei University. Her research interests encompass quantum algorithms, quantum machine learning, quantum applications in the field of PNT, and alternative PNT systems.
Ms. Jeong received the Undergraduate Fellowships from the Information and Communications Technology (ICT) Consilience Creative Program supported by the Ministry of Science and ICT, Republic of Korea. She was also selected for the Education and Training Program of the Quantum Information Research Support Center, funded by the Ministry of Science and ICT, Republic of Korea. From September 2024 to September 2025, she was a Visiting Researcher at the University of Waterloo, Waterloo, ON, Canada.
\end{IEEEbiography}

\begin{IEEEbiography}
[{\includegraphics[width=1in,height=1.25in,clip,keepaspectratio]{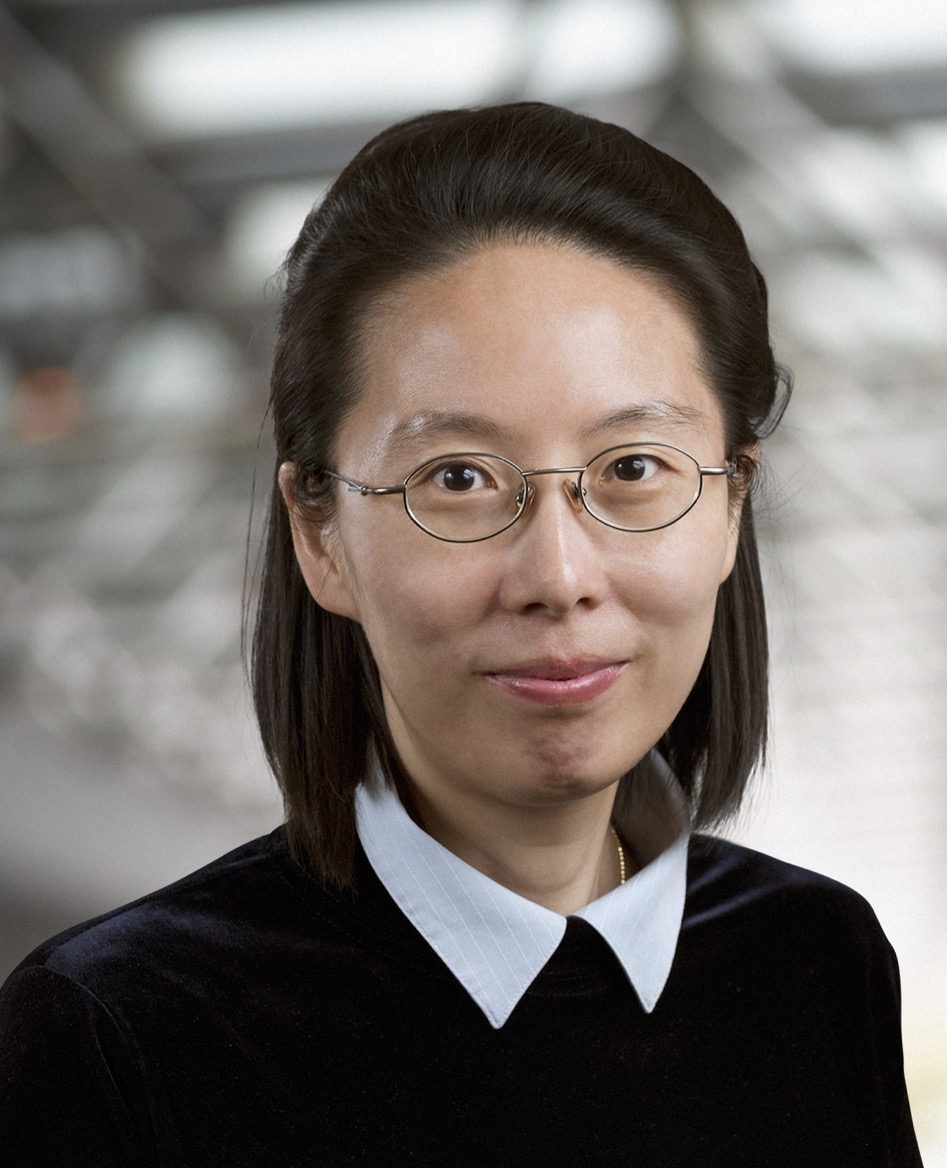}}]{Na Young Kim}
is a professor of electrical and computer engineering and a member of the Institute for Quantum Computing and the Waterloo Institute for Nanotechnology at the University of Waterloo in Canada. She received her doctoral degree in applied physics from Stanford University and a bachelor of science degree in physics from Seoul National University. Her research interests span quantum electronics, quantum optics, cavity quantum electrodynamics, condensed matter physics, and quantum information science and technology.
\end{IEEEbiography}

\begin{IEEEbiography}
[{\includegraphics[width=1in,height=1.25in,clip,keepaspectratio]{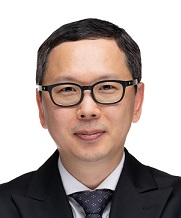}}]{Jiwon Seo} (Senior Member, IEEE) received the B.S. degree in mechanical engineering (division of aerospace engineering) from the Korea Advanced Institute of Science and Technology (KAIST), Daejeon, Republic of Korea, in 2002, the M.S. degree in aeronautics and astronautics, the second M.S. degree in electrical engineering, and the Ph.D. degree in aeronautics and astronautics from Stanford University, Stanford, CA, USA, in 2004, 2008, and 2010, respectively. 
He is currently an Underwood Distinguished Professor at Yonsei University, where he is a Professor in the School of Integrated Technology, Incheon, Republic of Korea. 
His research interests include GNSS anti-jamming technologies, complementary PNT systems, and intelligent unmanned systems. 
Dr. Seo is a member of the International Advisory Council of the Resilient Navigation and Timing Foundation, Alexandria, VA, USA, and Advisory Committee on Defense of the Presidential Advisory Council on Science and Technology, Republic of Korea.
\end{IEEEbiography}

\vfill

\end{document}


\title{\textit{Supplementary Material for}\\
Quantum-Based k-Coverage Optimization for UAV-Aided Search and Rescue Missions}

\author{Halim~Lee, Suhui~Jeong, Na~Young~Kim, and Jiwon~Seo,~\IEEEmembership{Senior Member,~IEEE}}

\maketitle

\begin{strip}
\noindent\textbf{Description:} This document provides supplementary details for the manuscript ``Quantum-Based k-Coverage Optimization for UAV-Aided Search and Rescue Missions.'' It provides formal proofs of the NP-hardness of the extended $k$-coverage problem and the exactness condition of the proposed QUBO reformulation. It further examines QAOA convergence, run-to-run stability, calibration-derived quantum-noise effects, and sensitivity to the principal QUBO and QAOA settings. In addition, it describes the implementation and training of the PPO-RL baseline and provides implementation details of the classical optimization baselines used in the comparative evaluation.
 
\vspace{1ex}
\noindent\rule{\textwidth}{0.4pt}
\vspace{0.5ex}
\end{strip}

\appendices

\section{NP-Hardness of the Extended $k$-Coverage Problem} \label{AppendixA}
\label{app:nphard}

In this appendix, we establish the computational complexity of the extended $k$-coverage problem introduced in Problem~2.

\begin{proposition}
\label{prop:nphard}
The extended $k$-coverage problem in Problem~2 is NP-hard.
\end{proposition}

\begin{proof}
Recall that the extended $k$-coverage problem is defined over a UAV-navigable region $\mathcal{U}$ and a target region $\mathcal{T}$, where the objective is to select a minimum-cardinality subset of candidate waypoints such that every grid point in $\mathcal{T}$ is covered by at least $k$ selected waypoints.

Consider the special case in which $\mathcal{U} = \mathcal{T}$. In this case, the extended $k$-coverage problem reduces exactly to the conventional $k$-coverage problem, since the candidate waypoint region and the target region coincide. Therefore, the conventional $k$-coverage problem is a special case of Problem~2.

Since the conventional $k$-coverage problem is NP-hard \cite{Hefeeda07:Randomized}, and Problem~2 strictly generalizes it, the extended $k$-coverage problem is also NP-hard.
\end{proof}

\begin{remark}
The above reduction also shows that any exact algorithm for the extended $k$-coverage problem would solve the conventional $k$-coverage problem as a special case.
\end{remark}

\section{Exactness of the QUBO Reformulation} \label{AppendixB}
\label{app:qubo_exactness}

In this appendix, we establish a sufficient exactness condition
for the canonical uniform-penalty QUBO formulation in
Eq.~(13).

Let the QUBO objective be written as
\begin{equation}
\label{eq:app_qubo_obj}
\mathcal{L}(\mathbf{x}, \mathbf{c}) = \sum_{i\in \boldsymbol{W}} x_i  + \lambda \cdot \sum_{j\in \boldsymbol{G}} \left( \sum_{i \in \boldsymbol{N}_j} x_i - k - \sum_{b=0}^{B_j-1} 2^b c_{j,b} \right)^2,
\end{equation}
where $x_i \in \{0,1\}$ indicates whether waypoint $i \in \boldsymbol{W}$ is selected, and the binary slack variables $c_{j,b} \in \{0,1\}$ encode the nonnegative excess coverage at grid point $j \in \boldsymbol{G}$.

\begin{proposition}
\label{prop:qubo_exactness}
Suppose that Problem~2 has at least
one feasible solution and that the slack-variable encoding can
represent the excess coverage associated with every feasible
solution. If $\lambda>|\boldsymbol{W}|$, then every global
minimizer of $\mathcal{L}(\mathbf{x},\mathbf{c})$ is feasible for
Problem~2. Moreover, for every
feasible $\mathbf{x}$,
\[
\min_{\mathbf{c}}\mathcal{L}(\mathbf{x},\mathbf{c})
=
\sum_{i\in\boldsymbol{W}}x_i.
\]
Therefore, minimizing $\mathcal{L}(\mathbf{x},\mathbf{c})$ is
equivalent to minimizing the original objective over the feasible
set.
\end{proposition}

\begin{proof}
We prove the result in two steps.

First, consider any feasible solution $\mathbf{x}$ of Problem~2. By feasibility, for every grid point $j \in \boldsymbol{G}$, we have
\[
\sum_{i \in \boldsymbol{N}_j} x_i \ge k.
\]
Hence, for each $j \in \boldsymbol{G}$, the nonnegative integer
\[
s_j = \sum_{i \in \boldsymbol{N}_j} x_i - k
\]
can be represented by the binary slack variables $\{c_{j,b}\}_{b=0}^{B_j-1}$, by assumption on the slack encoding. Therefore, the squared penalty term for each $j$ can be made exactly zero. Consequently, for every feasible solution, there exists a choice of slack variables such that
\[
\mathcal{L}(\mathbf{x},\mathbf{c}) = \sum_{i \in \boldsymbol{W}} x_i.
\]
Thus, over the feasible set, minimizing the QUBO objective is equivalent to minimizing the original cardinality objective.

Second, consider any infeasible assignment $\mathbf{x}$. Then there exists at least one grid point $j^\star \in \boldsymbol{G}$ such that
\[
\sum_{i \in \boldsymbol{N}_{j^\star}} x_i < k.
\]
For this grid point, no choice of nonnegative slack variables can satisfy
\[
\sum_{i \in \boldsymbol{N}_{j^\star}} x_i - k - \sum_{b=0}^{B_{j^\star}-1} 2^b c_{j^\star,b} = 0,
\]
because the left-hand side is strictly negative before introducing the slack term, while the slack term is nonnegative. Since all variables are binary, the squared residual for this violated constraint is at least $1$.
Therefore, for every infeasible $\mathbf{x}$ and every choice of
$\mathbf{c}$,
\[
\mathcal{L}(\mathbf{x},\mathbf{c})
\ge
\sum_{i\in\boldsymbol{W}}x_i+\lambda
\ge
\lambda
>
|\boldsymbol{W}|.
\]
On the other hand, because
Problem~2 is feasible, there exists a
feasible solution $\mathbf{x}^{\mathrm{F}}$ and a corresponding
slack assignment $\mathbf{c}^{\mathrm{F}}$ such that
\[
\mathcal{L}(\mathbf{x}^{\mathrm{F}},\mathbf{c}^{\mathrm{F}})
=
\sum_{i\in\boldsymbol{W}}x_i^{\mathrm{F}}
\le
|\boldsymbol{W}|.
\]
Hence, no infeasible assignment can be a global minimizer.
Furthermore, for each feasible $\mathbf{x}$, minimizing over
$\mathbf{c}$ eliminates all penalty terms. Thus, the global
minimizers of the QUBO correspond exactly to minimum-cardinality
feasible solutions of Problem~2.
This proves the claim.
\end{proof}

\section{QAOA Convergence, Stability, Noise Effects, and Parameter Sensitivity}
\label{app:qaoa_sensitivity}

The analyses in this appendix complement the feasibility and optimality evaluation in Section~IV by examining QAOA convergence, run-to-run stability, performance under calibration-derived quantum noise, and sensitivity to the principal QUBO and QAOA settings.
All experiments use scenario~(g), one of the eight test scenarios
shown in Fig.~4.
This instance comprises $|\boldsymbol{W}|=7$ candidate waypoints and $|\boldsymbol{G}|=6$ target grid points, and its known
minimum number of selected waypoints for full 3-coverage is five.
Using the same problem instance throughout these experiments enables controlled one-factor-at-a-time comparisons without confounding changes in the target geometry or QUBO structure.

Unless otherwise specified, the reference configuration used a QAOA depth of
$p=2$, 1024 shots per objective-function evaluation, constrained optimization
by linear approximations (COBYLA) as the classical optimizer, and a maximum
budget of one objective-function evaluation. In each sensitivity experiment,
only the factor under investigation was varied while all other settings were
fixed at their reference values. Except for the convergence trace in Section~\ref{supp:Convergence}, each configuration was executed ten times using the same set of random seeds, and continuous results are reported as the mean and sample standard deviation.

The primary application-level evaluation based on the solutions retained by the best-of-shots (BoS) rule includes the number of selected waypoints, mean 3-coverage ratio, feasible rate, and approximation ratio.
Except for the penalty-coefficient sweep, these metrics were invariant across all tested configurations and independent executions:
$5.00\pm0.00$ selected waypoints, a mean 3-coverage ratio of $100.00\%\pm0.00\%$, a feasible rate of $100\%$ $(10/10)$, and an
approximation ratio of $1.000\pm0.000$.
To avoid redundant reporting, these invariant metrics are omitted from the subsequent sensitivity tables, which focus on the quantities that vary with each experimental setting.

To characterize the full sampled distribution, we additionally report
the feasible and optimal probability masses, denoted by
$P_{\mathrm{feas}}$ and $P_{\mathrm{opt}}$, respectively.
Let $\mathcal{X}=\{0,1\}^{|\boldsymbol{W}|}$ denote the waypoint-decision
space, $\mathcal{F}\subseteq\mathcal{X}$ the set of decisions satisfying
all 3-coverage constraints, and $\mathcal{X}^{\star}\subseteq\mathcal{F}$ the set of globally minimum-cardinality feasible decisions.
For the $m$ measurement shots obtained from the final QAOA circuit, let $h(\mathbf{x})$ denote the number of shots whose waypoint-decision component is $\mathbf{x}$,
irrespective of the associated auxiliary slack-bit values.
Accordingly, $\sum_{\mathbf{x}\in\mathcal{X}}h(\mathbf{x})=m$.
The empirical probability masses are defined as
\begin{align}
P_{\mathrm{feas}}
&= \frac{1}{m}\sum_{\mathbf{x}\in\mathcal{F}}h(\mathbf{x}),
\label{eq:app-pfeas}\\
P_{\mathrm{opt}}
&= \frac{1}{m}\sum_{\mathbf{x}\in\mathcal{X}^{\star}}h(\mathbf{x}).
\label{eq:app-popt}
\end{align}
Thus, $P_{\mathrm{feas}}$ is the fraction of measured samples whose
waypoint decisions satisfy all 3-coverage constraints, whereas
$P_{\mathrm{opt}}$ is the fraction corresponding to globally
minimum-cardinality feasible decisions. All such decisions are included
when multiple global optima exist. Since
$\mathcal{X}^{\star}\subseteq\mathcal{F}$,
$0\leq P_{\mathrm{opt}}\leq P_{\mathrm{feas}}\leq1$.

These quantities complement the BoS metrics, which are computed from
one lowest-objective sample retained per run. In contrast,
$P_{\mathrm{feas}}$ and $P_{\mathrm{opt}}$ use all $m$ shots and should
be interpreted as per-shot probability masses rather than per-run BoS
success rates.

\subsection{Convergence Behavior}
\label{supp:Convergence}
To assess the convergence behavior of the QAOA parameter search, the maximum objective-function evaluation budget was increased from one to 70 while the remaining reference
settings were held fixed.
Although up to 70 evaluations were permitted, COBYLA satisfied its stopping criterion and terminated after 48 objective-function evaluations.
Therefore, the parameter search was regarded as converged according to the optimizer's stopping rule under the tested configuration.
The expected QUBO objective was recorded at every objective-function evaluation
through the QAOA callback.
This quantity is dimensionless and differs from the expected cost-Hamiltonian
value only by the constant offset introduced by the QUBO-to-Ising
transformation; hence, the two quantities follow the same optimization
trajectory.

\begin{figure}[t]
    \centering
    \IfFileExists{figures/Convergence.png}{%
        \includegraphics[width=0.7\columnwidth]{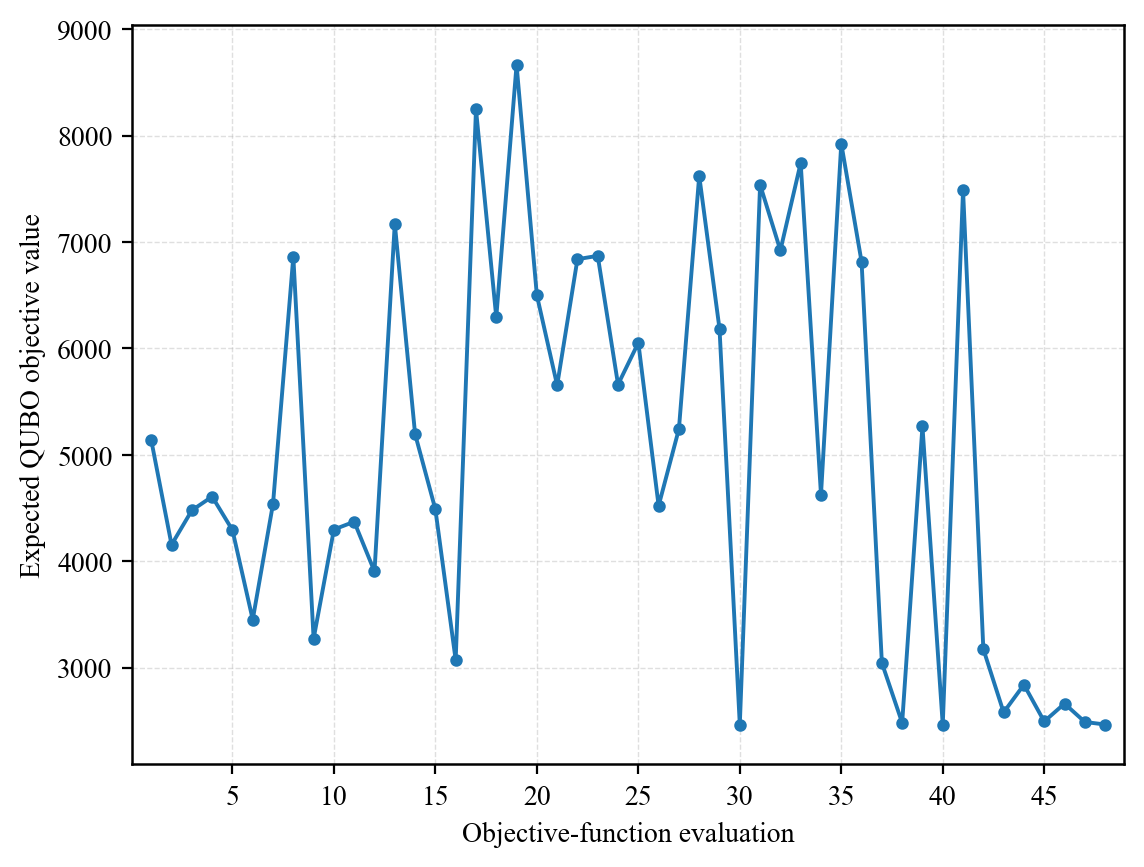}%
    }{%
        \fbox{\parbox[c][0.45\columnwidth][c]{0.94\columnwidth}{%
        \centering Insert \texttt{figures/Convergence.png}}}%
    }
    \caption{Objective-function evaluation history of the QAOA parameter
    search for scenario~(g). Each marker denotes the expected QUBO objective evaluated at a trial parameter vector. Although the callback history is nonmonotonic, the terminal evaluations occupy a lower and narrower objective range, and COBYLA satisfies its stopping criterion after 48 evaluations.}
    \label{fig:supp_convergence}
\end{figure}

As shown in Fig.~\ref{fig:supp_convergence}, the objective history is
nonmonotonic. COBYLA is a derivative-free trust-region optimizer that
constructs local linear approximations from sampled objective values and
evaluates exploratory trial points within the trust region. Because the
callback records all such evaluations rather than only the incumbent solution
retained by the optimizer, the recorded history need not be monotonic.
Although isolated low-objective values occur during the intermediate search and are followed by substantial increases, the final evaluations remain consistently within a lower and narrower objective range. Together with the satisfaction of COBYLA's stopping criterion at the 48th evaluation, this late-stage stabilization indicates convergence of the parameter search according to the optimizer's stopping rule.
This optimizer-level convergence does not by itself imply that the global QUBO minimum was reached.

\subsection{Stability Across Independent Executions}

Across ten independent executions of the reference configuration, every run
returned a feasible, globally optimal solution with five selected waypoints,
$100\%$ 3-coverage, and an approximation ratio of $1.000$; the run-to-run
standard deviation was zero for all three metrics. Based on the waypoint labels
used for scenario~(g) in the main manuscript, $\{2,3,4,6,7\}$ was obtained in
five executions, $\{1,2,4,6,7\}$ in three, and $\{1,3,4,5,7\}$ in two. All
three sets satisfy the six 3-coverage constraints and contain the known minimum
of five waypoints. The observed variation therefore represents selection among
objective-equivalent global optima rather than variation in solution quality.

\subsection{Robustness to Calibration-Derived Quantum Noise}
\label{supp:noise_robustness}
An approximate device-noise model was constructed from an
\texttt{ibm\_yonsei} calibration snapshot using Qiskit Aer. The same
physical-qubit subset, coupling map, and transpilation settings were retained
for all conditions. For a nominal local error channel
$\mathcal{E}_{g,\mathbf{q}}$ associated with gate $g$ and physical qubits
$\mathbf{q}$, we evaluated integer channel exposures
$\eta\in\{0,1,2,3\}$ according to
\begin{equation}
\mathcal{E}_{g,\mathbf{q}}^{(\eta)}
=
\mathcal{E}_{g,\mathbf{q}}^{\circ\eta},
\qquad
\mathcal{E}_{g,\mathbf{q}}^{(0)}=\mathcal{I}.
\label{eq:supp_noise_exposure}
\end{equation}
Thus, $\eta=0$ and $\eta=1$ denote ideal and nominal calibration-derived
conditions, respectively, whereas $\eta=2$ and $\eta=3$ represent controlled
twofold and threefold channel-exposure stress tests rather than exact
multiplication of the physical error probabilities.

\begin{table}[t]
\caption{Feasible- and optimal-solution probability masses under
calibration-derived quantum-noise exposure in scenario~(g). Values are reported
as the mean $\pm$ sample standard deviation over ten independent executions.}
\label{tab:supp_noise_robustness}
\centering
\footnotesize
\setlength{\tabcolsep}{9pt}
\begin{tabular}{ccc}
\toprule
\textbf{Exposure $\boldsymbol{\eta}$}
&
\boldmath{$P_{\mathrm{feas}}$} \textbf{(\%)}
&
\boldmath{$P_{\mathrm{opt}}$} \textbf{(\%)}
\\
\midrule
0 & $10.45 \pm 1.78$ & $3.35 \pm 0.68$ \\
1 & $9.14 \pm 1.12$  & $3.18 \pm 0.65$ \\
2 & $9.16 \pm 1.00$  & $3.03 \pm 0.60$ \\
3 & $8.97 \pm 1.06$  & $2.95 \pm 0.63$ \\
\bottomrule
\end{tabular}
\end{table}

Across all noise-exposure conditions, the best-of-shots results were identical: five selected waypoints, $100\%$ mean 3-coverage, a feasible rate of $100\%$, and an approximation ratio of $1.000$.
The run-to-run standard deviation was zero for all four metrics.
Quantum noise is generally expected to degrade solution quality; in this experiment, however, its effect was more apparent in the sampled distribution than in the solution retained by the BoS rule.
The BoS rule selects only one lowest-objective sample from the $m=1024$ shots, and a moderate redistribution of probability mass therefore need not alter the retained solution as long as favorable samples remain in the shot batch.
Even at $\eta=3$, the mean $P_{\mathrm{opt}}=2.95\%$ corresponds to approximately 30 shots per execution whose waypoint-decision components are globally optimal.
Although this waypoint-marginal probability is not equivalent to the joint QUBO ground-state probability, it shows that optimal waypoint decisions remained repeatedly represented in the sampled distribution. This persistence helps explain why the final BoS solution
remained unchanged despite the increased noise exposure.

The distribution-level results nevertheless suggest a modest degradation. From $\eta=0$ to $\eta=3$, $P_{\mathrm{feas}}$ decreased overall from $10.45\pm1.78\%$ to $8.97\pm1.06\%$, while $P_{\mathrm{opt}}$ decreased from $3.35\pm0.68\%$ to $2.95\pm0.63\%$. Because the intermediate response was nonmonotonic and the run-to-run variations overlapped, these results do not establish a strictly monotonic exposure--response relationship.
Moreover, the relatively low probability masses already observed under the ideal condition indicate that the limited concentration of the output distribution cannot be attributed to calibration-derived noise alone; finite QAOA depth, parameter optimization, and the QUBO landscape may also contribute.
The present experiment applies different channel exposures to a fixed circuit and therefore does not reproduce the additional error locations and circuit duration arising as the problem size and QAOA depth increase.
For larger and deeper QAOA circuits, cumulative gate errors and decoherence remain important constraints on scalability.
Accordingly, the unchanged BoS results demonstrate solution-level robustness only for scenario~(g) with 1024 shots, rather than general noise tolerance of QAOA.

\subsection{Hyperparameter Sensitivity}
\label{supp:hyperparameter_sensitivity}

We examined the effects of QAOA circuit depth, penalty coefficient,
objective-function evaluation budget, classical optimizer, and shot count.
Except for the penalty-coefficient sweep, the converter-generated QUBO used in
the noise-free validation was retained. For the penalty sweep, the
converter-selected penalties were replaced by a common scalar coefficient
$\lambda$ supplied to all constraint-conversion stages. This procedure enables
a controlled assessment of the canonical uniform-penalty formulation and the
sufficient exactness condition established in the main manuscript.
Tables~\ref{tab:supp_penalty_sensitivity} and
\ref{tab:supp_qaoa_hyperparameter_sensitivity} summarize the results.

\subsubsection{QAOA Layer Depth}

For scenario~(g), the QAOA layer depth was varied over
$p\in\{1,2,3\}$ under noise-free simulation, while all remaining
settings were held fixed. At every tested depth, the BoS rule retained
a globally optimal waypoint solution in every execution. Thus, the
application-level BoS quality had already saturated at $p=1$, with no
further improvement observed at higher depths.

Differences appeared at the distribution level. The mean
$P_{\mathrm{opt}}$ values were $3.12\%$, $3.21\%$, and $4.42\%$ for
$p=1$, $2$, and $3$, respectively. Hence, $p=3$ yielded the largest
mean probability mass on globally optimal waypoint decisions, with a
numerical increase of $1.30$ percentage points over $p=1$. This increase
did not improve the retained solution because BoS selects only one
lowest-objective sample and a global optimum was already retained in
every execution at $p=1$.

In the tested configuration, the mean $P_{\mathrm{feas}}$ varied
nonmonotonically with $p$. QAOA minimizes the expected QUBO objective,
which is an energy-weighted average over the joint waypoint--slack
distribution, rather than directly maximizing $P_{\mathrm{feas}}$ or
$P_{\mathrm{opt}}$. The standard depth-$(p+1)$ ansatz contains the
depth-$p$ ansatz, so its best attainable expected objective cannot
worsen under exact parameter optimization. This variational property,
however, does not guarantee monotonic improvement in either probability
mass. The observed variation should therefore be interpreted as a
distributional response specific to the tested scenario and settings,
rather than as evidence that increasing $p$ generally provides no
benefit.

Each additional layer also appends one cost and one mixer block. Under
noise-free simulation, this increased circuit size introduces neither
gate errors nor decoherence; accordingly, $p=3$ provided a higher mean
$P_{\mathrm{opt}}$ but no additional BoS-quality gain. On noisy
hardware, the additional gates and circuit duration may further
increase exposure to cumulative gate errors and decoherence, which is
not captured by the present experiment.

\subsubsection{Penalty Coefficient}
The bound $\lambda>|\boldsymbol{W}|$ provides an instance-independent sufficient condition for exactness of the uniform-penalty QUBO.
Because this condition is sufficient rather than necessary, a tighter instance-specific threshold was determined for scenario~(g) by exhaustive enumeration.

To determine the instance-specific threshold, the binary slack variables
in Eq.~(13) were minimized out. For a fixed waypoint
assignment $\mathbf{x}$, the resulting objective is
\begin{equation}
\label{eqn:reduced_QUBO}
\begin{aligned}
Q_{\lambda}(\mathbf{x})
&:= \min_{\mathbf{c}} \mathcal{L}(\mathbf{x},\mathbf{c}) \\
&= M(\mathbf{x})+\lambda P_{\min}(\mathbf{x}), \\
P_{\min}(\mathbf{x})
&:= \sum_{j\in\boldsymbol{G}}
\left[
\max\left\{0,\,k-\nu_j(\mathbf{x})\right\}
\right]^2,
\end{aligned}
\end{equation}
where
$M(\mathbf{x})=\sum_{i\in\boldsymbol{W}}x_i$ is the number of selected
waypoints and
$\nu_j(\mathbf{x})=\sum_{i\in\boldsymbol{N}_j}x_i$ is the coverage
multiplicity of target grid point $j$.

For scenario~(g), $k=3$, and the feasible optimum selects
$M^\star=5$ waypoints with zero penalty, yielding
$Q_{\lambda}(\mathbf{x}^{\star})=5$. For a lower-cardinality infeasible
assignment to have a strictly larger objective value than the feasible
optimum, the penalty coefficient must satisfy
\begin{equation}
\lambda>
\frac{M^\star-M(\mathbf{x})}{P_{\min}(\mathbf{x})}.
\end{equation}
Assignments with $M(\mathbf{x})\geq M^\star$ cannot improve upon the
feasible optimum for $\lambda>0$. The exact instance-specific threshold
is therefore
\begin{equation}
\label{eqn:lambda_critical}
\lambda_{\mathrm{crit}}
=
\max_{\substack{
\mathbf{x}\notin\mathcal{F}\\
M(\mathbf{x})<M^\star
}}
\frac{M^\star-M(\mathbf{x})}{P_{\min}(\mathbf{x})}.
\end{equation}

Exhaustive enumeration of all $2^{7}$ waypoint assignments shows that
the maximum is attained by a four-waypoint assignment in which two
target grid points have coverage multiplicity two and all remaining
targets satisfy the required 3-coverage. The two unit coverage deficits
give
\begin{equation}
P_{\min}(\mathbf{x})=1^2+1^2=2,
\qquad
Q_{\lambda}(\mathbf{x})=4+2\lambda.
\end{equation}
Consequently,
\begin{equation}
\lambda_{\mathrm{crit}}
=
\frac{5-4}{2}
=0.5.
\end{equation}
Therefore, $\lambda<0.5$ favors this infeasible assignment over the
feasible optimum, $\lambda=0.5$ permits a tie, and $\lambda>0.5$
ensures strict instance-level exactness.

\begin{table}[t]
\caption{Penalty-coefficient sensitivity for scenario~(g)
($\lambda_{\mathrm{crit}}=0.5$).}
\label{tab:supp_penalty_sensitivity}
\centering
\scriptsize
\setlength{\tabcolsep}{2.5pt}
\renewcommand{\arraystretch}{1.08}
\resizebox{\columnwidth}{!}{%
\begin{tabular}{@{}ccccc@{}}
\toprule
\boldmath{$\lambda$}
&
\makecell{\textbf{Mean 3-Cov.}\\\textbf{Ratio (\%)}}
&
\makecell{\textbf{Feasible}\\\textbf{Rate}}
&
\boldmath{$P_{\mathrm{feas}}$} \textbf{(\%)}
&
\boldmath{$P_{\mathrm{opt}}$} \textbf{(\%)}
\\
\midrule
$0.25$
& $33.33 \pm 11.11$
& $0\%$ $(0/10)$
& $18.06 \pm 12.42$
& $2.89 \pm 0.63$
\\
$0.50$
& $76.67 \pm 22.50$
& $40\%$ $(4/10)$
& $15.93 \pm 13.52$
& $3.52 \pm 1.06$
\\
$1.00$
& $100.00 \pm 0.00$
& $100\%$ $(10/10)$
& $11.95 \pm 6.21$
& $3.09 \pm 0.89$
\\
$|\boldsymbol{W}|+1=8$
& $100.00 \pm 0.00$
& $100\%$ $(10/10)$
& $9.97 \pm 2.27$
& $3.45 \pm 1.19$
\\
$2|\boldsymbol{W}|=14$
& $100.00 \pm 0.00$
& $100\%$ $(10/10)$
& $11.18 \pm 3.58$
& $3.67 \pm 1.06$
\\
\bottomrule
\end{tabular}
}
\end{table}

To evaluate penalty sensitivity around the instance-specific boundary, $\lambda$ was varied below, at, and above $\lambda_{\mathrm{crit}}=0.5$, with ten independent executions per setting and all remaining QAOA settings held fixed.
In Table~\ref{tab:supp_penalty_sensitivity}, the 3-coverage ratio, $P_{\mathrm{feas}}$, and $P_{\mathrm{opt}}$ are reported as the mean $\pm$ sample standard deviation, whereas the feasible rate denotes the fraction of executions whose BoS solution was feasible.

The results exhibit the predicted boundary behavior. At $\lambda=0.25$, boundary-defining infeasible assignments have lower QUBO values than the feasible optimum; for example, the four-waypoint assignment has $4+2\lambda=4.5<5$. Consequently, an infeasible BoS
solution was retained in every execution, yielding a feasible rate of $0\%$ and a mean 3-coverage ratio of $33.33\%$. The nonzero $P_{\mathrm{feas}}$ and $P_{\mathrm{opt}}$ at the same coefficient do not conflict with the zero feasible rate: these probabilities are
computed using all shots, whereas the feasible rate evaluates only the single BoS solution retained per execution. Feasible and globally optimal waypoint decisions could therefore be sampled without being retained when a lower-QUBO-cost infeasible state was also observed.
At $\lambda=0.5$, the feasible and boundary-defining infeasible objectives are tied, permitting either type of solution to be retained; the observed feasible rate was $40\%$. 
Every tested value above the boundary yielded feasible minimum-cardinality BoS solutions in all ten executions.
Further increases in $\lambda$ did not improve the application-level BoS results, while $P_{\mathrm{feas}}$ and $P_{\mathrm{opt}}$ varied nonmonotonically because exactness constrains the global minimizers of the QUBO rather than the entire output distribution produced by finite-depth QAOA.

\subsubsection{Objective-Function Evaluation Budget}
The objective-function evaluation budget for COBYLA, controlled by
\texttt{maxiter}, was varied over $1$, $10$, $30$, and $50$, while all
remaining settings were held fixed. For every tested budget and every
independent execution, the BoS rule retained a globally optimal,
minimum-cardinality waypoint solution satisfying full 3-coverage.
Accordingly, the application-level BoS performance remained unchanged
across the tested budgets, whereas differences were observed in the
sampled output distributions.

The mean $P_{\mathrm{feas}}$ increased monotonically from
$10.23\% \pm 2.03\%$ at one evaluation to
$14.41\% \pm 11.58\%$, $18.90\% \pm 22.89\%$, and
$19.74\% \pm 23.08\%$ at budgets of $10$, $30$, and $50$,
respectively. Although the mean probability increased by $9.51$
percentage points over the tested range, the concurrent growth in the
sample standard deviation indicates increasingly heterogeneous
outcomes across executions. Because each objective value was estimated
from 1024 measurement shots, small sampling differences could lead
independent executions to follow different COBYLA search paths.
COBYLA constructs local linear approximations and evaluates exploratory
trial points within an adaptive trust region; therefore, the sequence
of recorded objective evaluations is not required to decrease
monotonically. This behavior is also visible in the convergence
analysis, where the intermediate evaluations fluctuate before entering
a stable terminal region. Since COBYLA minimizes the expected QUBO
objective rather than $P_{\mathrm{feas}}$ directly, additional
evaluations can produce substantially greater feasible-state
concentration in some executions without yielding comparable gains in
others. The observed mean increase from $30$ to $50$ evaluations was
only $0.84$ percentage points, suggesting diminishing returns over this
interval despite the additional evaluation cost.

The mean $P_{\mathrm{opt}}$ likewise varied nonmonotonically, taking
values of $3.21\%$, $2.93\%$, $3.36\%$, and $3.98\%$ as the budget
increased. This behavior is consistent with the convergence analysis,
as COBYLA minimizes the expected QUBO objective rather than
$P_{\mathrm{opt}}$ directly. Nevertheless, the mean
$P_{\mathrm{opt}}$ remained within a narrow range of
$2.93\%$--$3.98\%$, corresponding to approximately $30$--$41$
globally optimal samples per 1024-shot final distribution. The
per-execution BoS results further show that at least one globally
optimal decision satisfying full 3-coverage was sampled and retained
in every execution.

\subsubsection{Classical Optimizer}
COBYLA and simultaneous perturbation stochastic approximation (SPSA) were
compared using the same budget of 30 objective-function evaluations. In
contrast to the derivative-free trust-region COBYLA described in
Section~\ref{supp:Convergence}, SPSA approximates a stochastic
gradient from symmetric simultaneous perturbations. The comparison was
controlled by the actual number of objective-function evaluations rather than
by optimizer-specific stopping parameters because COBYLA and SPSA interpret
their iteration-control settings differently.

The mean $P_{\mathrm{opt}}$ values were nearly identical, at
$3.36\pm0.78\%$ for COBYLA and $3.35\pm0.89\%$ for SPSA. COBYLA produced a
higher mean $P_{\mathrm{feas}}$ but substantially greater run-to-run dispersion
($18.90\pm22.89\%$ versus $12.88\pm4.03\%$). Under the fixed SPSA gain and
perturbation settings, SPSA therefore yielded more consistent feasible-state
probability across executions, whereas COBYLA reached higher feasible-state
concentration in a subset of runs. The two optimizers nevertheless showed no
discernible difference in optimal-state probability under the matched
evaluation budget.

\subsubsection{Number of Shots}
Changing the shot count from 256 to 4096 had little effect on the mean
output-probability metrics. The mean $P_{\mathrm{feas}}$ remained between
$10.16\%$ and $10.32\%$, while the mean $P_{\mathrm{opt}}$ remained between
$3.21\%$ and $3.52\%$. In contrast, the run-to-run dispersion decreased with
increasing shot count: the standard deviation of $P_{\mathrm{feas}}$ decreased
from $2.46$ to $1.61$ percentage points, and that of $P_{\mathrm{opt}}$
decreased from $1.43$ to $0.65$ percentage points.

This reduction is consistent with lower finite-sampling uncertainty, for which
the standard error of a fixed output probability scales approximately as
$S^{-1/2}$ under independent sampling. Since the mean probability masses were
already stable at 256 shots and the additional reduction in
$P_{\mathrm{opt}}$ dispersion beyond 1024 shots was limited, 1024 shots were
retained as a practical compromise between probability-estimation variability
and sampling cost.

\begin{table}[t]
\caption{Output-probability sensitivity to QAOA, sampling, and classical optimization
settings. Values are reported as the mean $\pm$ sample standard deviation over
ten independent executions.}
\label{tab:supp_qaoa_hyperparameter_sensitivity}
\centering
\scriptsize
\setlength{\tabcolsep}{6pt}
\renewcommand{\arraystretch}{1.05}
\begin{tabular}{lcc}
\toprule
\textbf{Setting}
&
\boldmath{$P_{\mathrm{feas}}$} \textbf{(\%)}
&
\boldmath{$P_{\mathrm{opt}}$} \textbf{(\%)}
\\
\midrule
\multicolumn{3}{l}{\textit{QAOA circuit depth}}\\
$p=1$ & $12.73\pm5.28$ & $3.12\pm1.22$ \\
$p=2$ & $10.23\pm2.03$ & $3.21\pm0.75$ \\
$p=3$ & $13.98\pm2.53$ & $4.42\pm1.25$ \\

\addlinespace[2pt]
\multicolumn{3}{l}{\textit{Objective-function evaluation budget}}\\
$1$  & $10.23\pm2.03$  & $3.21\pm0.75$ \\
$10$ & $14.41\pm11.58$ & $2.93\pm1.16$ \\
$30$ & $18.90\pm22.89$ & $3.36\pm0.78$ \\
$50$ & $19.74\pm23.08$ & $3.98\pm1.47$ \\

\addlinespace[2pt]
\multicolumn{3}{l}{\textit{Classical optimizer
(30 objective-function evaluations)}}\\
COBYLA & $18.90\pm22.89$ & $3.36\pm0.78$ \\
SPSA   & $12.88\pm4.03$  & $3.35\pm0.89$ \\

\addlinespace[2pt]
\multicolumn{3}{l}{\textit{Number of shots}}\\
$256$  & $10.16\pm2.46$ & $3.52\pm1.43$ \\
$1024$ & $10.23\pm2.03$ & $3.21\pm0.75$ \\
$4096$ & $10.32\pm1.61$ & $3.36\pm0.65$ \\
\bottomrule
\end{tabular}
\end{table}

\color{black}
\section{PPO-RL Baseline Implementation}
\label{AppendixC}
 
This appendix summarizes the MDP formulation, network architecture, training procedure, and cross-domain and in-domain training regimes of the PPO-RL baseline evaluated in the main manuscript.
 
\subsection{MDP Formulation}
 
The waypoint selection task is formulated as a finite-horizon sequential decision-making problem in which the agent iteratively selects candidate waypoints to add to the solution set and issues a STOP action to terminate the episode.

\subsubsection{State}
The state vector $\mathbf{s}_t \in \mathbb{R}^{5W_{\max} + 5G_{\max} + 4}$ concatenates three feature groups:
\begin{equation}
    \mathbf{s}_t = \bigl[\,\mathbf{w}_1, \ldots, \mathbf{w}_{W_{\max}},\;
                          \mathbf{g}_1, \ldots, \mathbf{g}_{G_{\max}},\;
                          \mathbf{f}\,\bigr],
\end{equation}
where $W_{\max}$ and $G_{\max}$ denote the maximum number of candidate waypoints and target points observed across all training instances, respectively, and slots beyond $|\boldsymbol{W}|$ and $|\boldsymbol{G}|$ are zero-padded.
Each waypoint feature vector is $\mathbf{w}_i = [\,\tilde{x}_i,\, \tilde{y}_i,\, s_i,\, \phi_i,\, v_i\,]$, with $(\tilde{x}_i, \tilde{y}_i)$ denoting normalized spatial coordinates, $s_i \in \{0,1\}$ the selection status, $C_i^{\mathrm{sat}} = \{j : p_j \in \boldsymbol{C}_i,\, \mathrm{cnt}_j < k\}$ the set of currently under-covered target points that waypoint $i$ can newly satisfy, $\phi_i = |C_i^{\mathrm{sat}}| / |\boldsymbol{G}|$ the normalized marginal coverage gain, and $v_i \in \{0,1\}$ the slot validity indicator (i.e., $v_i = 1$ if slot $i$ corresponds to an actual waypoint, and $v_i = 0$ for padding slots).
Each target feature vector is $\mathbf{g}_j = [\,\tilde{x}_j,\, \tilde{y}_j,\, d_j,\, c_j,\, v_j\,]$, where $\mathrm{cnt}_j$ denotes the number of currently selected waypoints covering target point $p_j$, $d_j = \max(0,\, k - \mathrm{cnt}_j) / k$ is the normalized coverage deficit, $c_j = \mathrm{cnt}_j / k$ is the normalized coverage count, and $v_j \in \{0,1\}$ is the slot validity indicator for target slots.
The global feature vector is $\mathbf{f} = [\,|\boldsymbol{W}|/W_{\max},\; |\boldsymbol{G}|/G_{\max},\; t/W_{\max},\; \rho_k\,]$, where $t$ is the current step count and $\rho_k = |\{j : \mathrm{cnt}_j \geq k\}| / |\boldsymbol{G}|$ is the fraction of target points already satisfying the $k$-coverage requirement.
 
\subsubsection{Action}
The action space consists of $W_{\max} + 1$ discrete actions: selecting waypoint $i \in \{0, \ldots, |\boldsymbol{W}|-1\}$, or issuing a STOP action ($a = W_{\max}$).
Invalid actions (already-selected waypoints and padded slots) are suppressed via action masking~\cite{Huang22:Closer}, which sets the logit of invalid actions to $-\infty$ before the softmax, ensuring the agent never selects an invalid action.
 
\subsubsection{Reward}
The per-step reward for selecting waypoint $i$ is
\begin{equation}
    r_t = -1 + \alpha \cdot \Delta_{\mathrm{sat}},
\end{equation}
where
\begin{equation}
    \Delta_{\mathrm{sat}} = \sum_{j \in G}
        \Bigl[\min\!\bigl(\mathrm{cnt}_j^{\mathrm{new}},\, k\bigr)
             - \min\!\bigl(\mathrm{cnt}_j^{\mathrm{old}},\, k\bigr)\Bigr]
\end{equation}
is the incremental gain in satisfied coverage after adding waypoint $i$, capped at $k$ per target to avoid over-rewarding redundant coverage, and $\alpha = 0.5$ is a weighting coefficient.
Upon termination, a bonus of $+50$ is awarded if the solution is feasible (i.e., $\mathrm{cnt}_j \geq k$ for all $j \in \boldsymbol{G}$), and a penalty of $-100$ is applied otherwise.
 
\subsection{Network Architecture, Hyperparameters, and Training}

Two training regimes were considered while retaining the same MDP formulation, network architecture, and PPO hyperparameters.

\begin{figure}[tbp]
    \centering
    \includegraphics[width=0.7\linewidth]{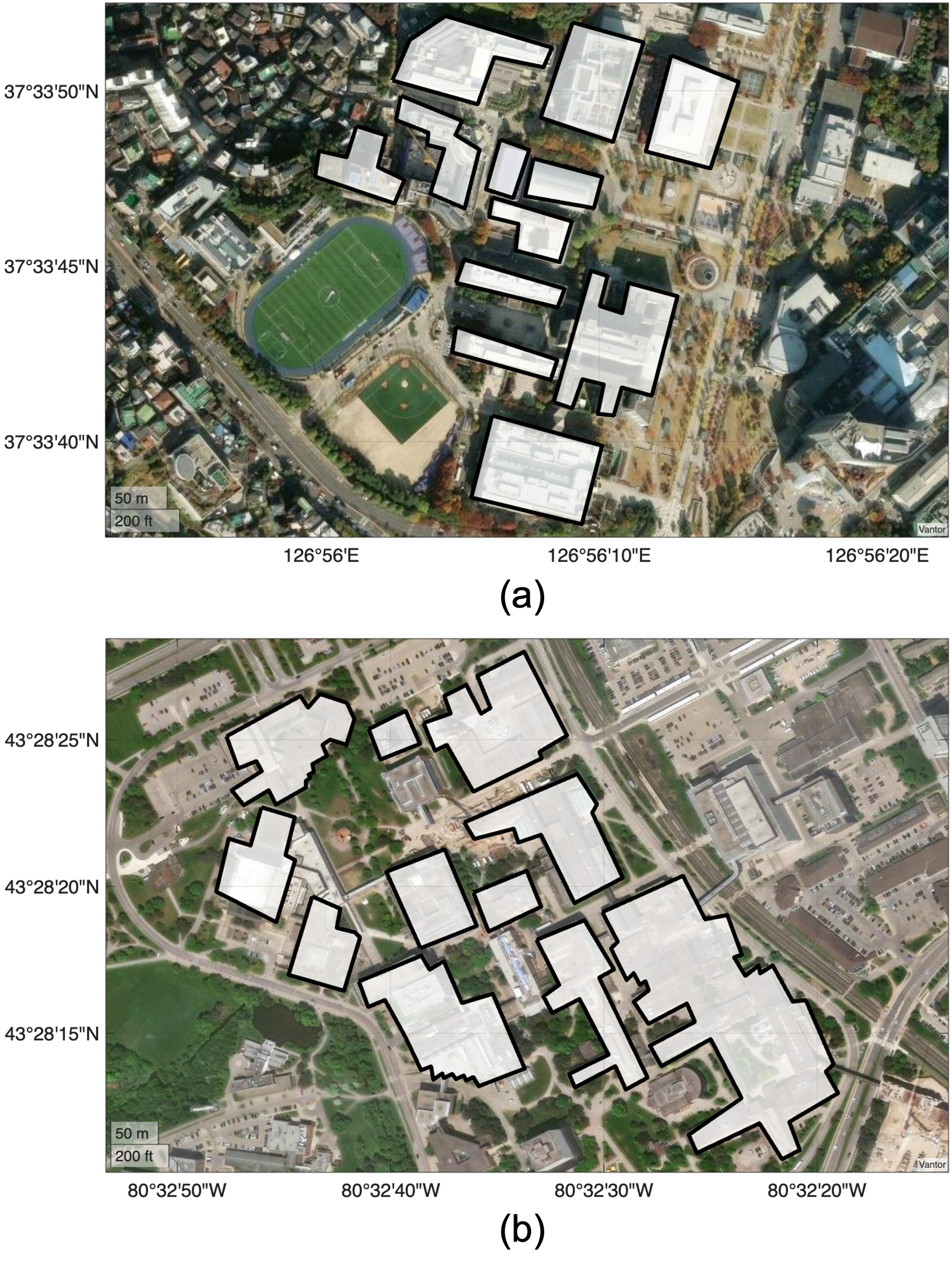}
    \caption{
Satellite images of the two campuses used for PPO-RL training.
(a)~Yonsei University Seoul Campus, South Korea.
(b)~University of Waterloo, Canada.
Buildings used as target areas are highlighted with semi-transparent white regions and black boundaries.
The test campus (Yonsei University International Campus) was geographically distinct and entirely unseen during training.
}
    \label{fig:training_campuses}
\end{figure}

For the \emph{cross-domain} regime, the agent was trained for 200{,}000 steps on 23 building-area configurations from Yonsei University Seoul Campus and the University of Waterloo, with grid spacing, communication range, and UAV altitude randomized across instances to promote generalization.
The evaluation scenario at Yonsei University International Campus was entirely unseen during training.
Satellite images of the two campuses used for cross-domain training are shown in Fig.~\ref{fig:training_campuses}.

For the \emph{in-domain} regime, the same PPO architecture and hyperparameters were retained, but the exact evaluation instance from Yonsei University International Campus was included in the
training distribution.
Accordingly, the in-domain result is not a held-out generalization result; rather, it is used as a favorable upper-bound reference to assess the effect of environment-specific training on PPO-RL performance.

\begin{table}[ht]
\centering
\caption{Key PPO-RL hyperparameters.}
\label{tab:ppo_hyperparams}
\begin{tabular}{lc}
\toprule
\textbf{Hyperparameter} & \textbf{Value} \\
\midrule
Algorithm               & MaskablePPO~\cite{Huang22:Closer} \\
Policy / value network  & MLP [256, 256] \\
Learning rate           & $3 \times 10^{-4}$ \\
Batch size              & 256 \\
Discount factor & 0.99 \\
Total training steps    & 200{,}000 \\
\bottomrule
\end{tabular}
\end{table}
 
\subsection{Training Convergence}
 
\begin{figure}[tbp]
    \centering
    \includegraphics[width=0.9\linewidth]{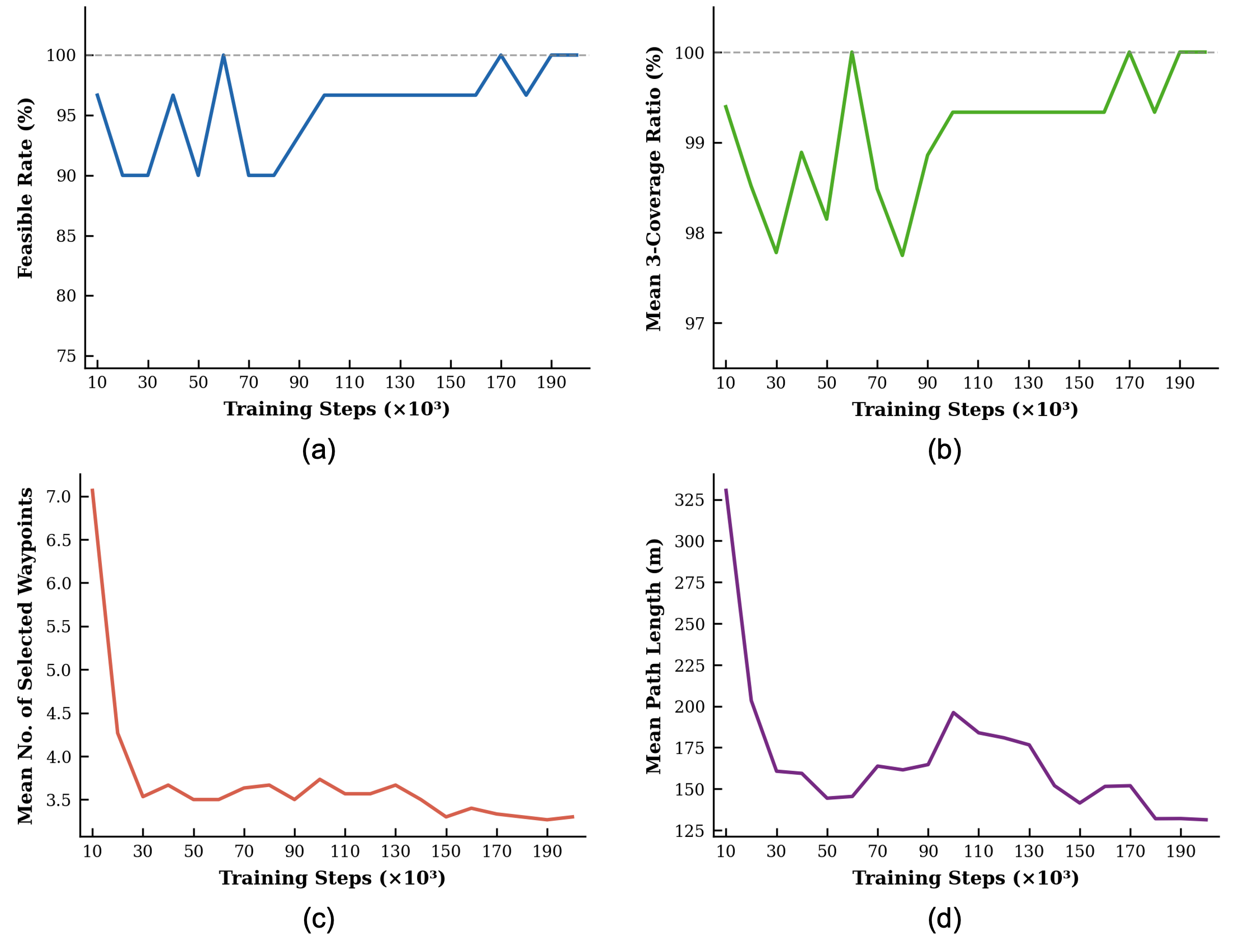}
    \caption{
Validation metrics during PPO-RL training evaluated every 10{,}000 steps:
(a)~feasibility rate, (b)~mean 3-coverage ratio,
(c)~mean number of selected waypoints, and (d)~mean path length.
Both feasibility rate and mean 3-coverage ratio reach 100\% by 200{,}000 steps.
}
    \label{fig:validation_curves}
\end{figure}

For the cross-domain training regime, Fig.~\ref{fig:validation_curves} shows four validation metrics evaluated every 10{,}000 steps on held-out instances.
The feasibility rate (Fig.~\ref{fig:validation_curves}(a)) rises rapidly from 96.7\% at 10{,}000 steps and reaches 100\% by 200{,}000 steps, indicating that the agent quickly learns to produce solutions that fully satisfy the 3-coverage requirement.
The mean 3-coverage ratio (Fig.~\ref{fig:validation_curves}(b)) similarly converges to 100\%, confirming that coverage is achieved consistently across all target points rather than on average only.
 
The mean number of selected waypoints (Fig.~\ref{fig:validation_curves}(c)) decreases from 7.1 at 10{,}000 steps to approximately 3.3 at convergence, reflecting the agent's progressive learning to select a more compact waypoint subset.
The mean path length (Fig.~\ref{fig:validation_curves}(d)) decreases correspondingly from 330~m to approximately 131~m, consistent with the reduction in waypoint count.
These results show that, under the cross-domain regime, the learned policy produces feasible and increasingly compact solutions on the held-out validation instances before evaluation on the geographically distinct test scenario.

\section{Classical Baseline Implementations}
\label{app:baselines}

This appendix details the classical baselines compared against the
proposed framework in Section~V-C. All
baselines operate on the same candidate waypoint set $\boldsymbol{W}$,
target set $\boldsymbol{G}$, and coverage sets $\{\boldsymbol{N}_j\}$
produced by the shared preprocessing stage, so that differences in
performance stem only from the optimization method. The coverage relation
is encoded as a sparse matrix
$\boldsymbol{A}_{\mathrm{cov}}\in\{0,1\}^{|\boldsymbol{G}|\times|\boldsymbol{W}|}$
with $(\boldsymbol{A}_{\mathrm{cov}})_{j,i}=1$ iff $i\in\boldsymbol{N}_j$.

\subsection{Exact ILP Baseline}
\label{app:ilp}

The exact baseline solves the constrained formulation in
Eq.~(9) directly as a binary integer linear program,
\begin{equation}
\label{eq:app_ilp}
\min_{\mathbf{x}\in\{0,1\}^{|\boldsymbol{W}|}}
\sum_{i\in\boldsymbol{W}} x_i
\quad\text{s.t.}\quad
-\boldsymbol{A}_{\mathrm{cov}}\mathbf{x} \le -k\mathbf{1},
\end{equation}
implemented with MATLAB's \texttt{intlinprog}, with all
$|\boldsymbol{W}|$ variables declared integer
($\texttt{intcon}=1{:}|\boldsymbol{W}|$) and box-constrained to
$[0,1]$. A time limit of 600~s is used for the campus scenario and
3600~s for the larger scaling instances. Unlike solvers that expose an
explicit optimality-gap target, \texttt{intlinprog} is run with its
default relative tolerance ($10^{-4}$); since the objective is
integer-valued, the search converges to a provably optimal integer
solution, whose cardinality we use as the optimality reference for the
approximation ratio reported in
Section~V-C. As the ILP encodes the original
constraints directly, it introduces no slack variables and is independent
of the QUBO formulation.

\subsection{Greedy Set Multi-Cover}
\label{app:greedy}

The greedy baseline builds the waypoint set incrementally. Let
$\mathrm{cnt}_j$ be the current coverage count of target $j$ and
$u_j=\mathbf{1}[\mathrm{cnt}_j<k]$ its deficit indicator. At each step, the gain of a candidate waypoint $i$ is $g_i=\sum_{j}(\boldsymbol{A}_{\mathrm{cov}})_{j,i}\,u_j$, i.e., the number of still-under-covered targets it would newly serve; the waypoint with the largest gain is added, with ties broken by lowest index, and the process repeats until every target is $k$-covered. A subsequent prune step then sorts the selected waypoints by ascending coverage size and removes any whose deletion still leaves all targets $k$-covered, eliminating redundant selections. Because the greedy construction already terminates at a feasible solution, only this prune step is exercised in practice. The algorithm is deterministic and is therefore executed once.

\subsection{Simulated Annealing on the QUBO Objective}
\label{app:sa}

The SA baseline follows a standard Metropolis-based annealing procedure~\cite{Kirkpatrick83:Optimization} and optimizes the QUBO formulation in Eq.~(13) over the full binary vector $\boldsymbol{z}\in\{0,1\}^{n}$, comprising the $|\boldsymbol{W}|$ waypoint-decision bits and $\sum_{j\in\boldsymbol{G}} B_j$ slack bits ($n=98$ for the campus scenario).
The penalty coefficient is fixed at $\lambda=2|\boldsymbol{W}|=78$, which satisfies the sufficient exactness condition $\lambda>|\boldsymbol{W}|$ established in Appendix~B.

At each move, one uniformly selected decision or slack bit is flipped to generate a neighboring solution. A move with QUBO objective change $\Delta L$ is accepted with probability $\min\{1,\exp(-\Delta L/T)\}$.
The temperature is decreased geometrically according to $T\leftarrow\alpha T$ with $\alpha=0.95$, starting from $T_0=\lambda$. At each temperature, $3n=294$ single-bit moves are evaluated. The annealing process terminates when the temperature falls below $T_f=10^{-3}$ or when the maximum of 300 temperature steps is reached, whichever occurs first.
SA is executed for ten independent runs with distinct random seeds under the same stochastic evaluation protocol used for
QAOA and PPO-RL.

\subsection{Genetic Algorithm on the QUBO Objective}
\label{app:ga}

The genetic algorithm (GA) baseline uses a binary population representation with standard selection, crossover, mutation, and elitism operators~\cite{Goldberg89:Genetic}.
It optimizes the same QUBO formulation as the SA baseline over $\boldsymbol{z}\in\{0,1\}^{n}$, with fitness given by $L(\boldsymbol{z})$ to be minimized.

A population of 200 individuals is evolved for 300 generations using binary tournament selection with tournament size two, uniform crossover with probability 0.8, and independent bit-flip mutation with probability $1/n$ per bit. One elite individual is retained unchanged in each generation. The initial population consists of one greedy-derived feasible solution, including its corresponding slack-bit assignment, and 199 uniformly generated random binary solutions. The GA uses $\lambda=2|\boldsymbol{W}|=78$ and is executed for ten independent random seeds under the same best-of-10 evaluation protocol.

\color{black}